\documentclass{emulateapj}
\usepackage{graphicx}
\usepackage[space]{grffile}
\usepackage{latexsym}
\usepackage{textcomp}
\usepackage{longtable}
\usepackage{multirow,booktabs}
\usepackage{amsfonts,amsmath,amssymb}
\usepackage{natbib,verbatim}
\usepackage[dvipsnames]{xcolor}
\usepackage{url}
\usepackage{hyperref}
\hypersetup{colorlinks=true,linkcolor=red,urlcolor=purple,citecolor=blue,pdfborder={0 0 0}}

\newif\iflatexml\latexmlfalse
\usepackage[utf8]{inputenc}
\usepackage[english]{babel}
\usepackage{todonotes}
\usepackage{xspace}
\usepackage{tabularx}
\usepackage{diagbox}
\usepackage{hhline}
\usepackage{comment}

\newcommand{\msun}{\,\mathrm{M}_\odot}

\newcommand{\pc}{\,\mathrm{pc}}
\newcommand{\au}{\,\mathrm{au}}

\newcommand{\aI}{a_{\rm I}}
\newcommand{\aO}{a_{\rm O}}

\newcommand{\abh}{a_{\rm SMBH}}
\newcommand{\abin}{a_{\rm bin}}
\newcommand{\Rh}{R_{\rm Hill}}
\newcommand{\kms}{\,\mathrm{km}/\mathrm{s}}
\newcommand{\rev}[1]{{#1}} %
\newcommand{\revi}[1]{{#1}} 

\shorttitle{The Keplerian three-body encounter}
\shortauthors{Trani et al.}

\definecolor{myred}{HTML}{cc3600}
\definecolor{myblue}{HTML}{3333cc}
\definecolor{mygreen}{HTML}{008000}

\begin{document}
	
	
	\title{The Keplerian three-body encounter II.\\ Comparisons with isolated encounters and impact on gravitational wave merger timescales}	
	
	
	\author{Alessandro A. Trani\altaffilmark{1,$\star$,$\dagger$}}
	\author{Mario Spera\altaffilmark{4,5,6}}
	\author{Nathan W.C. Leigh\altaffilmark{2,3}}
	\author{Michiko S. Fujii\altaffilmark{1}}

	\altaffiltext{1}{Department of Astronomy, Graduate School of Science, The University of Tokyo, 7-3-1 Hongo, Bunkyo-ku, Tokyo, 113-0033, Japan}
	\altaffiltext{2}{Departamento de Astronom\'ia, Facultad de Ciencias F\'isicas y Matem\'aticas, Universidad de Concepci\'on, Concepci\'on, Chile}
	\altaffiltext{3}{Department of Astrophysics, American Museum of Natural History, New York, NY 10024, USA}
	\altaffiltext{4}{INFN, Sezione di Padova, Via Marzolo 8, I--35131, Padova, Italy}
	\altaffiltext{5}{Center for Interdisciplinary Exploration and Research in Astrophysics (CIERA), Evanston, IL 60208, USA}
	\altaffiltext{6}{Department of Physics \& Astronomy, Northwestern University, Evanston, IL 60208, USA}
	\altaffiltext{$\star$}{Email: aatrani@gmail.com}
	\altaffiltext{$\dagger$}{JSPS Fellow}
	
	
	
	
	\begin{abstract}
	We investigate the role of the Keplerian tidal field generated by a supermassive black hole (SMBH) on the three-body dynamics of stellar mass black holes. We consider two scenarios occurring close to the SMBH: the breakup of unstable triples and three-body encounters between a binary and a single. \rev{These two cases correspond to the hard and soft binary cases, respectively.} The tidal field affects the breakup of triples by tidally limiting the system, so that the triples break earlier with lower breakup velocity, leaving behind slightly larger binaries (relative to the isolated case). The breakup direction becomes anisotropic and tends to follow the shape of the Hill region of the triple, favouring breakups in the radial direction. Furthermore, the tidal field can torque the system, leading to angular momentum exchanges between the triple and its orbit around the SMBH. This process changes the properties of the final binary, depending on the initial angular momentum of the triple. 
	Finally, the tidal field also affects binary-single encounters: binaries tend to become both harder and more eccentric with respect to encounters that occur in isolation.  Consequently, single-binary scattering in a deep Keplerian potential produces binaries with shorter gravitational wave merger timescales.
	\end{abstract}%

	\bibliographystyle{apj}
	
	
	
	
	
	\makeatletter
	\renewcommand\@dotsep{10000}
	\makeatother
	
	
	\section{Introduction}
	\begin{table*}[hbpt]
		\begin{center}
			\caption{\footnotesize Initial setup of bound triple systems.\label{tab:ictri}}
			\begin{tabular}{lcccccccc}
				\hhline{=========}
				Properties \textbackslash{} Set &	L	 & L1 	  & L2     & L3      & 	H	 & H1		& H2	 & H3 		\\ \hline
				$w$ 							&	0	 & 0  	  & 0  	   & 0  	 &	5	 & 5  	& 5 	 & 5  		\\
				$a_{\rm SMBH}$ [pc]				&  no SMBH	 & 0.1 	  & 0.01   & 0.005   & 	no SMBH	 & 0.1 	& 0.01   & 0.005  	\\
				$a_{\rm I}$	[au] 				&	1	 & 1  	  & 1 	   & 1       & 	1	 & 1  	& 1 	 & 1 		\\
				$a_{\rm O}$ [au] 				&	2	 & 2 	  & 2 	   & 2 	     & 	2	 & 2	 	& 2 	 & 2 		\\
				\hline
			\end{tabular}
		\end{center}
		\scriptsize 
		{\rev Row~1: dimensionless angular momentum $w$;}
		row~2: semimajor axis of the orbit around the SMBH; 
		row~3: semimajor axis of the inner binary of the triple; 
		row~4: semimajor axis of the outer binary of the triple.
	\end{table*}
	
	\begin{table*}[hbpt]
		\begin{center}
			\caption{\footnotesize Initial setup of single-binary encounters.\label{tab:icenc}}
			\begin{tabular}{lcccccccc}
				\hhline{=========}
				Properties \textbackslash{} Set & E1 		& E1p   & E1r & \revi{E1is}   	 & E2	     & E2p    & E2r  	 & E2is \\ \hline
				$\abh$ [pc] 	& $0.0291$ 	& $0.0291$ & $0.0291$ &  no SMBH      & $0.0291$ & $0.0291$ & $0.0291$ & no SMBH 	\\
				$i_\mathrm{bin}$  			& random 	& 0 	   & $\pi$    & random	 & random   & 0     	& $\pi$    & random    		\\
				$\abin$ [au] 		& 50 		& 50 	   & 50 	  & 50 		 & 5 	     & 5        & 5 	   &   5       		\\	
				$\abin/r_\mathrm{Hill}$& 1 		& 1 	   & 1 		  & 1 		 & 0.1 	 & 0.1      & 0.1 	   &       0.1    		\\
				
				$e_\mathrm{bin}$ 			& 0 		& 0 	   & 0 		  & 0		 & 0 	     & 0 	    & 0        &  0 	  		\\		
				$e_{\rm sin}$ 			& 0 		& 0 	   & 0 		  & 0		 & 0 	     & 0 	    & 0  	   &  0   \\ \hline
			\end{tabular}		
		\end{center}
		{\scriptsize 
			Row~1: semimajor axis of the binary orbit around the SMBH; 
			row~2: inclination of the inner binary orbit with respect to the orbital plane around the SMBH; 
			\revi{row~3: semimajor axis of the inner binary; 
			row~4: ratio between $a_{\rm in}$ and the Hill radius of the binary;}
			row~5: eccentricity of the inner binary; 
			row~6: mass of the binary components; 
			row~7: semimajor axis of the single body in orbit around the SMBH; 
			row~8: eccentricity of the single body in orbit around the SMBH.
		}
	\end{table*}
	Galactic nuclei are among the most dense environments in the Universe, where supermassive black holes (SMBH), giant molecular clouds and massive star clusters can coexist within a few parsecs.
	Compact remnants may sink towards the SMBH, forming an invisible yet extremely dense cusp \citep{bah76}. One such cusp may exist even in our Galactic center, as suggested by the X-ray binary population observed within one parsec \citep{hail18}.
	
	Binaries in galactic nuclei can encounter remnants from the cusp on short timescales. \citet{lei16a} calculated the rate of single-binary (1+2) interactions in Local Group nuclear star clusters and showed that, in spite of Keplerian motions near any SMBH and even for very low binary fractions $< 1\%$, the 1+2 encounter rate can be comparable to what is expected in globular clusters.  \citet{lei18} used an analytic Monte Carlo approach to study single-binary scatterings in galactic nuclei and active galactic nuclei (AGN) disks.  The authors showed that planar scatterings, as would occur in a disk, harden black hole binaries more efficiently relative to isotropic scatterings.  But they caution that the magnitude of the Coriolis force is typically comparable to the local gravitational force within the binary Hill sphere, and that more work is needed to better understand how the non-negligible Coriolis force affects the outcomes of single-binary interactions.
	
	In AGN disks, three-body scatterings between BH-BH binaries and other stars and/or compact objects are thought to occur commonly as well. \citet{secunda18} recently simulated the migration of stellar mass BHs in an analytically modeled AGN disk using an augmented N-body code.  Their simulations included migration torques, a stochastic gravitational force exerted by turbulent density fluctuations in the disk, damping of eccentricities and inclinations due to passages through the gas disk, and of course the usual gravitational forces exerted between objects. The authors found that BH-BH binaries can form rapidly and efficiently in AGN disks as the BHs migrate towards migration traps in their simulations. 
	
	We term the three-body encounters in which all encountering bodies are in orbit around the SMBH as Keplerian three-body encounters.
	\citet{tra2019} investigated Keplerian three-body encounters as a mechanism to produce S-stars and G-objects in the Galactic center via ionizing encounters between young stellar binaries and stellar mass black holes.
	Here we investigate the fundamental difference between isolated and Keplerian three-body encounters, by comparing our numerical experiments to the statistical escape theory of three-body breakups in isolation \citep{kart05}.
	We then use these results to estimate the impact of binary-single encounters in Keplerian potentials dominated by a central SMBH on the gravitational wave coalescence timescales of binary black holes.
	
\rev{
	In Section~\ref{sec:methods} we describe the numerical setup of our few-body simulations. Section~\ref{sec:triple} presents our results on the breakup of unstable hierarchical triples around SMBHs. In Section~\ref{sec:binsin} we present a detailed comparison between the isolated and Keplerian binary-single encounters, and in Section~\ref{sec:gw} we discuss the implications for the mergers of binary black holes in galactic nuclei. Finally, we summarize our results in Section~\ref{sec:conclusions}.
}

	\section{Numerical setup}\label{sec:methods}
	
	\rev{
	In order to assess the role of the Keplerian potential during three-body encounters, we examine two scenarios that cover the two extrema of spectrum of the velocity dispersion between the binary and the single: unstable hierarchical triples systems (low velocity dispersion) and unbound binary-single encounters (high velocity dispersion). 
	
	Another way to look at these two scenarios is the hard-soft binary limit \citep{heg75}. A binary is considered hard if its binding energy $E_\mathrm{b}$ is larger than the kinetic energy $E_\mathrm{k}$ of the encountering body. Conversely, if $E_\mathrm{b} < E_\mathrm{k}$, the binary is considered soft.
	In the unstable triple case, the outer body has a low relative velocity with respect to the inner binary, thus the inner binary is hard. In the case of binary-single hyperbolic interactions, the \revi{ relative velocity exceeds the orbital velocity} around the SMBH, so that, in the setup considered here, the binary is soft. 
	
	In the first case, we simulate unstable bound triples \revi{of point particles} in isolation. These triples decay into a binary and a single object, unbound with respect to each other. We then re-simulate the triples in different orbits around the SMBH and compare the outcomes with the isolated case.
	
	In the second scenario, we compare the outcomes of the same hyperbolic encounters between a binary and a single star both in a Keplerian potential and in isolation. 
		
	In all simulations, The SMBH mass is set to $4.31\times10^6 \msun$ \citep{gil09a,gil17}. All the other particles are equal mass black holes of $30 \msun$.
	In order to detect \revi{particle-particle collisions}, we set the radius of each particle to their Schwarzschild radius.
    \revi{Each set of simulations comprises $10^5$ individual realizations.}
}
	
	\subsection{Unstable hierarchical triples}
	Unstable triple systems inevitably decay in a finite time into an unbound binary and a single star (assuming all point-particles). The distribution of breakup velocities is well known in the context of statistical escape theory \citep{mon76a,mon76b,nas78,mik86,valt05,kart05}.
	The velocity distribution is dependent not only on the total energy of the initial system, but also on the total angular momentum of the triple \citep{anos69,stan72,sasl74,valt74,anos84,anos86,mik86,valt05,kart05,arc18a}. 
	In order to characterize the role of the Keplerian potential, we compare the breakup velocity vector for the same triple systems in isolation and in orbit around the SMBH.

	To generate unstable triples, we follow the procedure described in \citet{mik86}. The systems are initialized as unstable hierarchical triples. We fix the semimajor axis of the inner and outer orbit, and randomly generate the inner and outer eccentricity so that i) the \citet{mard01} criterion for hierarchical triple stability is not satisfied ii) the total angular momentum $L$ of the system satisfies the following relation:
	\begin{equation}
	w = \frac{-L^2 E_\mathrm{tot}}{G^2\, M_\mathrm{tot}^5}
	\end{equation}
	where $E_\mathrm{tot}$ is the total energy of the triple system and $w$ is a chosen dimensionless parameter that describes the amount of angular momentum in the system. For $w=0$, the total angular momentum of the triple is zero, i.e. the inner and outer orbits are coplanar and retrograde, so that their angular momentum is equal and anti-aligned. Since the system generated in this way is only approximately hierarchical, the initial conditions still show some residual angular momentum, leading to an effective $w_\mathrm{eff}\simeq0.01$ instead of zero. \rev{ We therefore refer to the initial conditions with $w=0$ as the low angular momentum case (L sets of simulations). We also simulate a set with $w = 5$, corresponding to the high angular momentum case (H sets of simulations). 
	
	We evolve the triple systems both in isolation and in circular orbits around the SMBH. \revi{In both the L and H sets}, once we set the triple in orbit around the SMBH, we randomize the direction of the total angular momentum of the triple, so that there is no preferred direction with respect to the orbital plane around the SMBH. In fact, the orientation of the total angular momentum vector of the triple is known to affect the angular distribution of breakup directions. Therefore, by randomizing the total angular momentum of the triple there will be no preferential breakup direction and any anisotropy can be due only to the coupling with the Keplerian tidal field. We set the semimajor axis of the inner and outer binary to $\aI = 1 \au$ \revi{and $\aO = 2 \au$}, respectively. With this configuration, the semimajor axis of the triple around the SMBH can be as small as $\abh = 0.005 \pc$ before the apocenter of the outer body ends up outside the Hill region of the triple.
	 Table~\ref{tab:ictri} lists the main initial conditions for the simulations of unstable hierarchical triple systems.
}

	\rev{
	The simulations are run for $10000\,t_\mathrm{dyn}$, where $t_\mathrm{dyn} = G M_\mathrm{tot}^{5/2}/ 2\left| E_\mathrm{tot}\right|^{3/2}$ is the dynamical time of the triple, or until the triple breaks up, whichever is shorter. As we will show in Section~\ref{sec:lifetime}, less then $0.1\%$ of the triples are still bound after $10^4 t_\mathrm{dyn}$. 
	}
	
	It is straightforward to obtain the breakup velocity of isolated three-body system by analyzing the final state of the system, consisting of the binary and the single body set on an escaping hyperbolic trajectory.
	It is not as simple when the three-body system breaks-up while on a Keplerian orbit around the SMBH. 
	In this case, after the triple breaks up, the single and the binary will be brought onto different orbits around the SMBH (or even become unbound from it). Given the orbital parameters of the triple and of the final objects (binary and single), it is possible to recover an analytic expression for the breakup velocity by assuming that the breakup velocity lies in the orbital plane of the triple. However, this assumption holds only if the original and final orbit lie in the same plane, and there is no analytic expression for the general case in which the inclination is altered.

	\rev{
	For this reason, we compute the breakup velocity as the difference between the velocity vector of the original orbit and the final orbit where their orbits cross (i.e. \revi{where the triple breaks up}).
	This requires knowing the true anomalies of both the original and final orbits at the crossing point. To get the true anomalies at the crossing point, we compute the minimum orbital intersecting distance (MOID\footnote{The MOID is defined as the distance between the closest points of two Keplerian orbits with a common focus. It is widely employed for the identification of potentially hazardous objects, i.e. comets and asteroids with a risk of impacting Earth. The original algorithm that we employed for the computation of the MOID is available at the following link: \url{http://moid.cbk.waw.pl/}.}) between the orbit of the triple and the orbits of the single and the binary
	using the method described in \citet{wizn18}, modified to work for arbitrary scales and hyperbolic orbits. For all cases we obtain a MOID of less than $1 \au$ ($\ll a_{\rm SMBH}$), indicating that the initial and final orbit do cross and that the impulsive approximation holds to a good approximation. The total breakup velocity is then the sum of the breakup velocity of the single and the binary.
}	
	
	\subsection{Binary-single encounters}
\rev{
	In order to set up the 4-body simulations in which a binary and a single object undergo a 3-body encounter while orbiting an SMBH, we follow the same method described in \citet{tra2019}. The orbits of the binary and the single around the SMBH are set up to be almost intersecting except for a small impact parameter, randomly sampled between 0 and twice the binary semimajor axis $\abin$. Earlier tests have shown that for \revi{an impact parameter} larger than $2\abin$ there is almost no interaction between the binary and the \revi{single \citep{leigh16c}}, and the shape of the distribution of impact parameters within this range \revi{has little} impact on the final outcome.
	The mutual inclination between the two orbits is chosen to be isotropic, consistent with the spherical symmetry and isotropy of \revi{the cusp around the SMBH in the Milky Way}. \revi{The simulations begin} with the binary and the single moving towards the orbital intersection, about $1/16$th of the binary orbital period before the encounter takes place. This is enough to ensure that the binary and the single are sufficiently separated \revi{at} the beginning of the simulation.
	
	If the binary breaks up during the encounter, we classify it as a ionization; if otherwise the original binary remains bound, we classify this outcome as \revi{a flyby}. If a member of the original binary is exchanged with the third body, we consider it an exchange.
	If a binary, either original or exchanged, survives the encounter, we wait for its center of mass to be 100 binary semimajor \revi{axes away} from the single body and record its orbital parameters. However, it may occur that the semimajor axis of the binary \revi{is positive} (hence indicating a bound binary) even when larger than the Hill radius of the system by orders of magnitude. In this case, the binary is in a temporary, loosely bound state but will \revi{eventually be broken up} by the tidal field. For this reason, after recording the binary orbital parameters we remove the third body and continue the simulation for 10 binary orbits around the SMBH. If the binary is still bound at the end of the simulation, we count the binary as bound, otherwise we consider it as \revi{an ionization}.	
	
 	Both the binary and the single \revi{orbit the SMBH on a circular orbit with $\abh=0.291\pc$}. We consider two cases: loose binaries with $\abin = 50 \au \simeq \Rh$, and tight binaries with $\abin = 5 \au \simeq 0.1 \Rh$. For the Hill radius, we employ the following definition:
 }
 	\begin{equation}\label{eq:hill}
 	r_\mathrm{Hill} = \abh \left(\frac{m_\mathrm{bin}}{3 M_\mathrm{SMBH}}\right)^{1/3}
 	\end{equation}
\rev{In both cases, the relative velocity between the binary and the single is much higher than the orbital velocity of the binary, so the binary can be considered as soft.  We also consider different binary inclinations with respect to the orbital plane around the SMBH and simulate three different sets for each binary semimajor axis considering prograde binaries, retrograde binaries and isotropic binary orientations.

	We re-run sets of Keplerian three-body encounters in isolation. In this case, we model the isolated hyperbolic encounters using the known impact parameter and relative velocity of the binary and single stars at the encounter from the simulations of Keplerian encounters. We set the relative velocity between the orbit of the single and the binary at the point of closest approach in the Keplerian simulations as the velocity at infinity in the isolated case. The impact parameter is the same as in the Keplerian encounters. Since in isolation there is no preferential reference frame, we re-run in isolation only the set with \revi{isotropic orientations} of the binaries.  

	Table~\ref{tab:ictri} lists the main initial conditions for our binary-single numerical scattering experiments.
}

	\subsection{\texttt{TSUNAMI} code}
	We run the simulations with \texttt{TSUNAMI}, a direct N-body integrator that implements Mikkola's algorithmic regularization, namely, the logarithmic Hamiltonian and the time-transformed leapfrog \citep{mik99a,mik99b}. When combined with a Bulirsh-Stoer extrapolation algorithm \citep{sto80} or a higher order symplectic scheme \citep{yoshi90}, this method is ideal to numerically integrate strong dynamical interactions with high mass ratios.	More details on the code will be presented in a following work (Trani et. al, in preparation).
	
	\texttt{TSUNAMI} includes velocity-dependent forces following the algorithm described in \citet{mik06,mik08}. Among these, we have implemented the post-Newtonian terms 1PN, 2PN and 2.5PN \citep{blanchet2006} and the tidal drag-force from \citet{sam18a}. Our code also includes collision detection.
	
	To better compare the results of the simulations with the (Newtonian) statistical theory of (isolated) three-body encounters, in the present work we do not turn on post-Newtonian terms. Post-Newtonian simulations with more realistic initial conditions will be presented in the next work of the series.
	
	
	\section{Unstable hierarchical triples around supermassive black holes}\label{sec:triple}
	
		\begin{figure}[!b]
		\includegraphics[width=\linewidth]{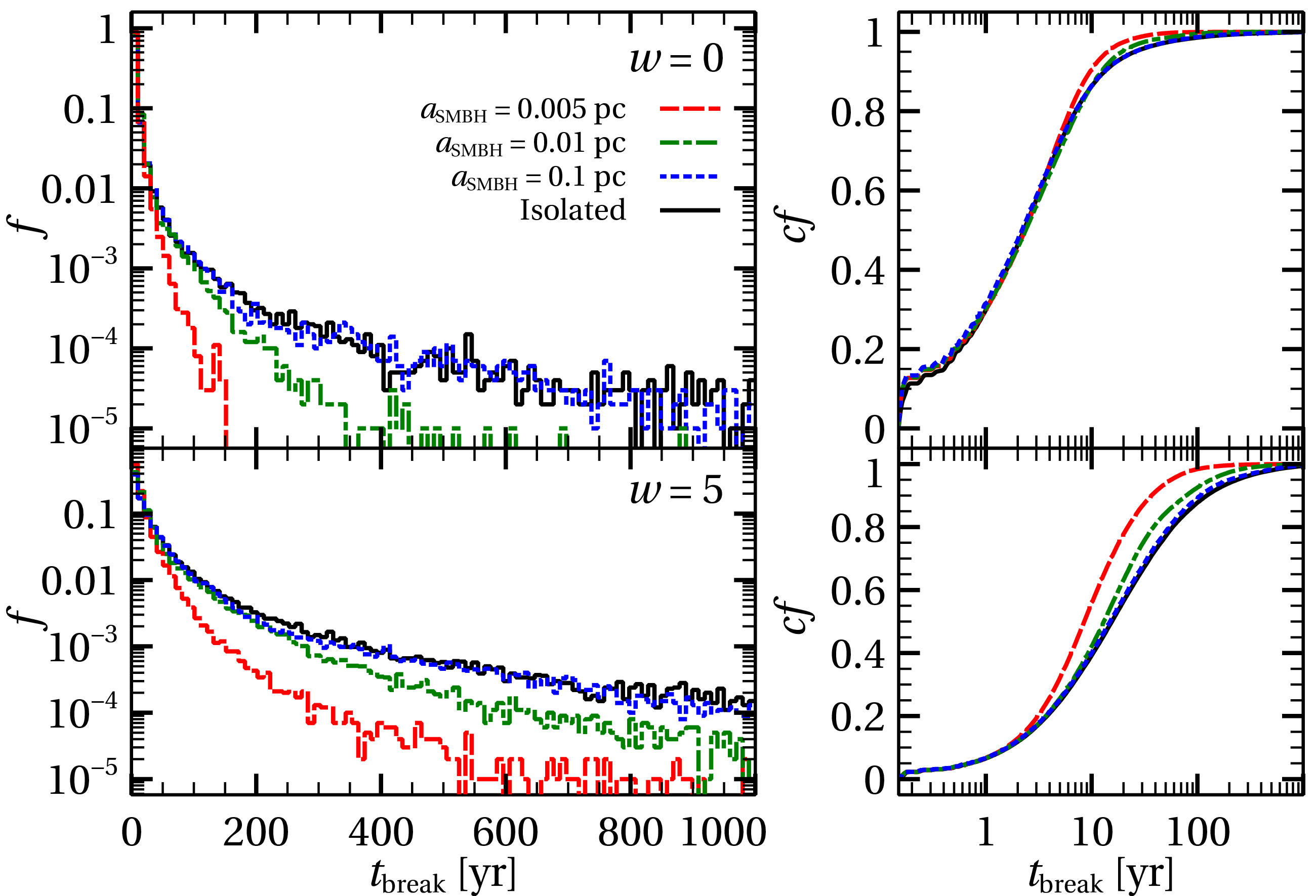}
		\caption{\footnotesize
			Fractional (left) and cumulative (right) distributions of the lifetimes of unstable triple systems. Dotted blue line: $a_{\rm SMBH} = 0.1\pc$; dot-dashed green line: $\abh = 0.01\pc$; dashed red line: $\abh = 0.01\pc$. Top panels: sets L, L1, L2 and L3 ($w = 0$). Bottom panel: sets H, H1, H2 and H3 ($w = 5$). The cumulative distributions are normalized to integrate to unity although each set has a different total number of broken up systems, mainly due to the different number of collided systems (especially in sets L, L1, L2 and L3; see Table~\ref{tab:outtri}).
		}
		\label{fig:tbreak}
	\end{figure}

	\subsection{Lifetime}\label{sec:lifetime}
	
	\begin{figure}[hbtp]
		\centering
		\includegraphics[width=0.9\linewidth]{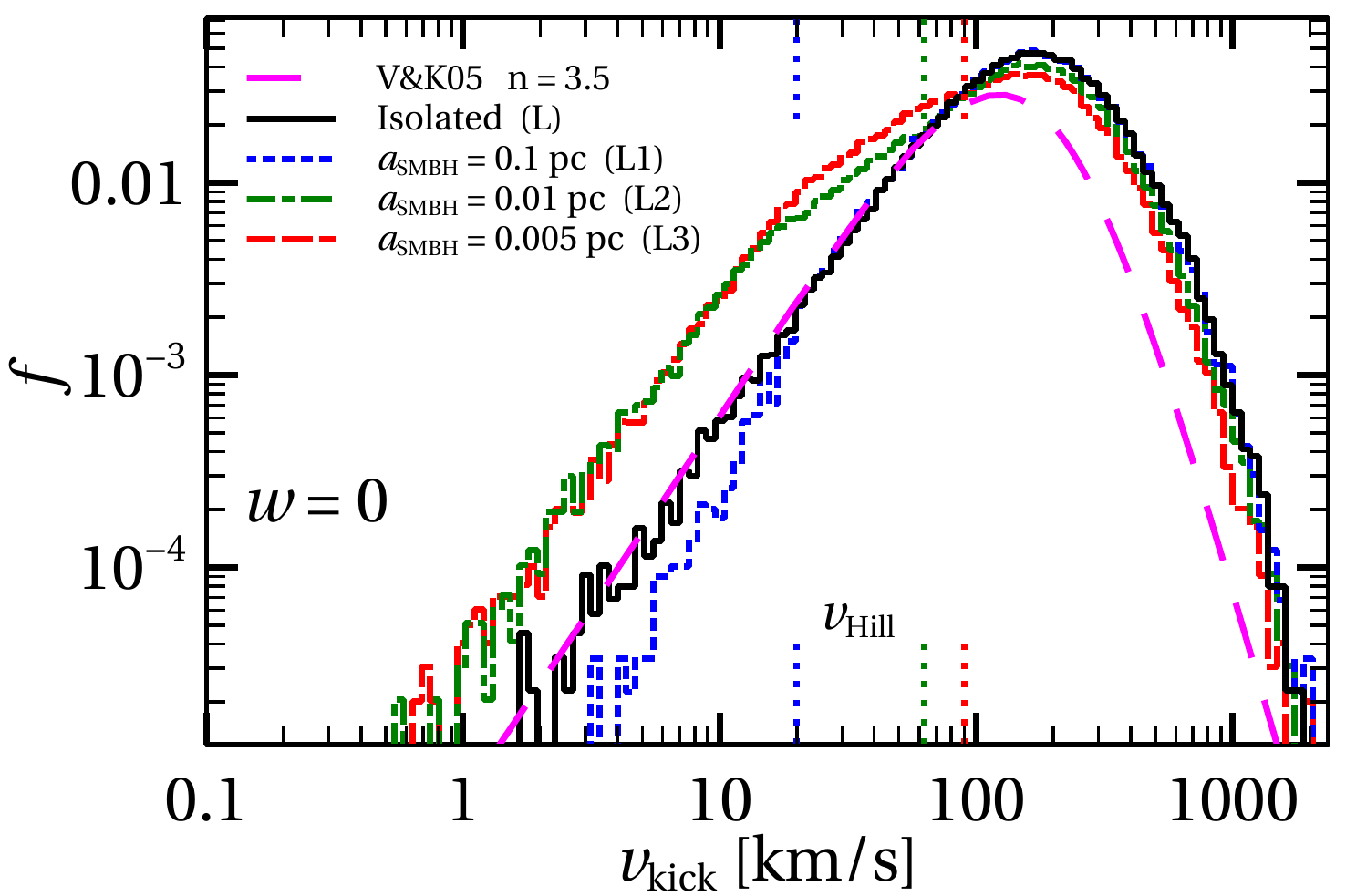}
		\includegraphics[width=0.9\linewidth,trim={1.6cm 1.9cm 1.3cm 1.9cm}, clip]{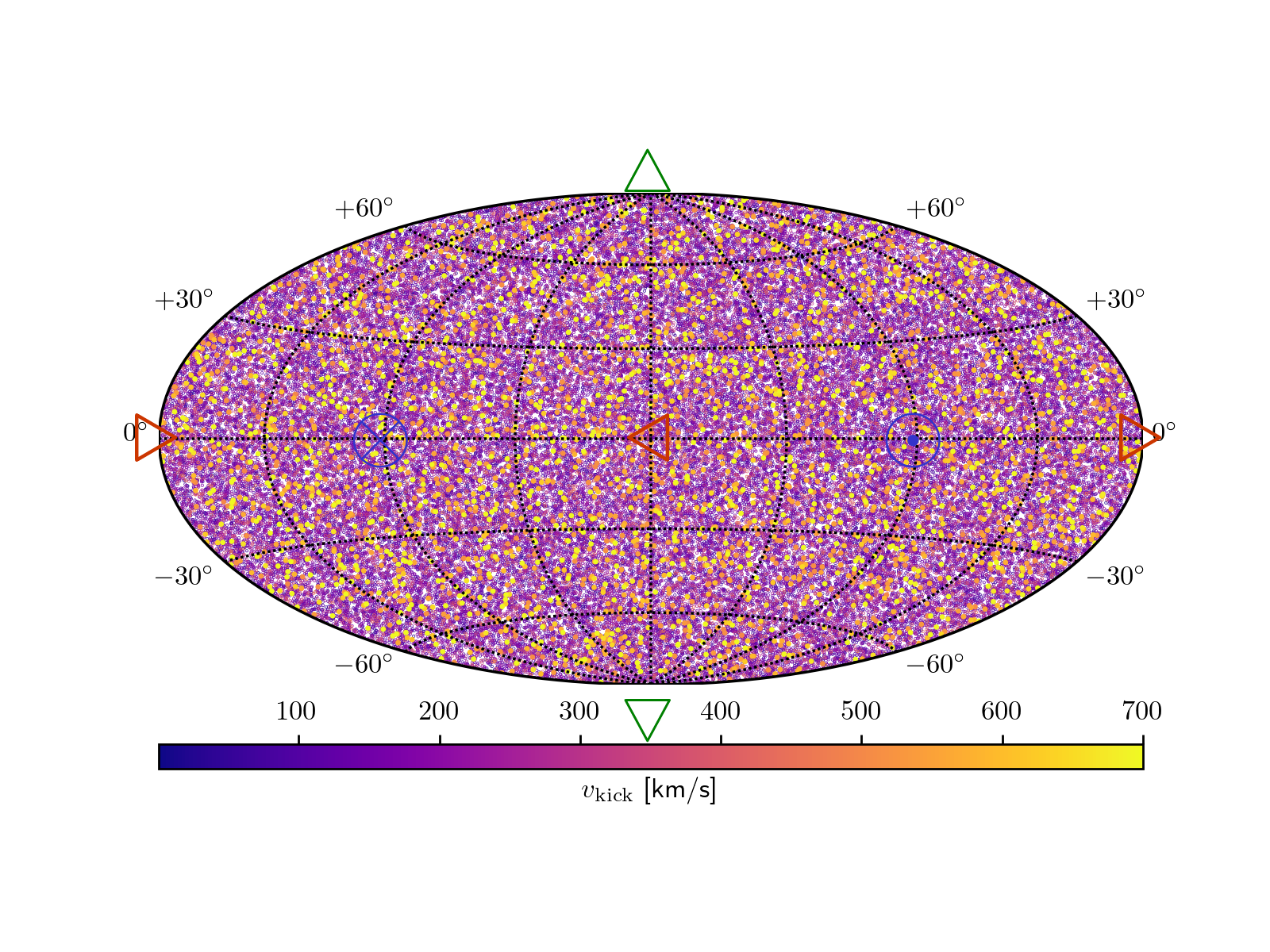}\\
		\includegraphics[width=0.9\linewidth,trim={1.6cm 1.9cm 1.3cm 1.9cm}, clip]{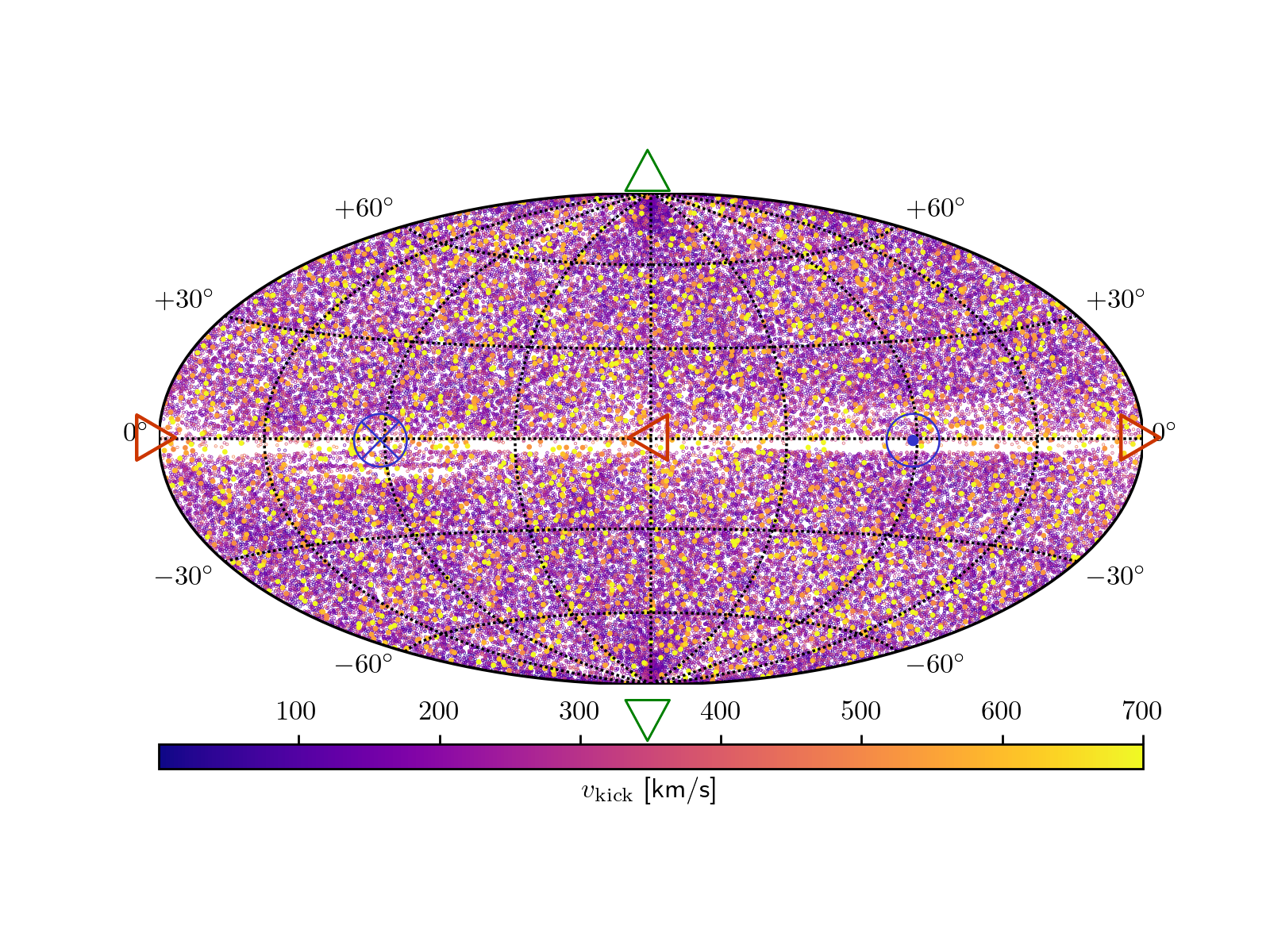}\\
		\includegraphics[width=0.9\linewidth,trim={1.6cm 1.9cm 1.3cm 1.9cm}, clip]{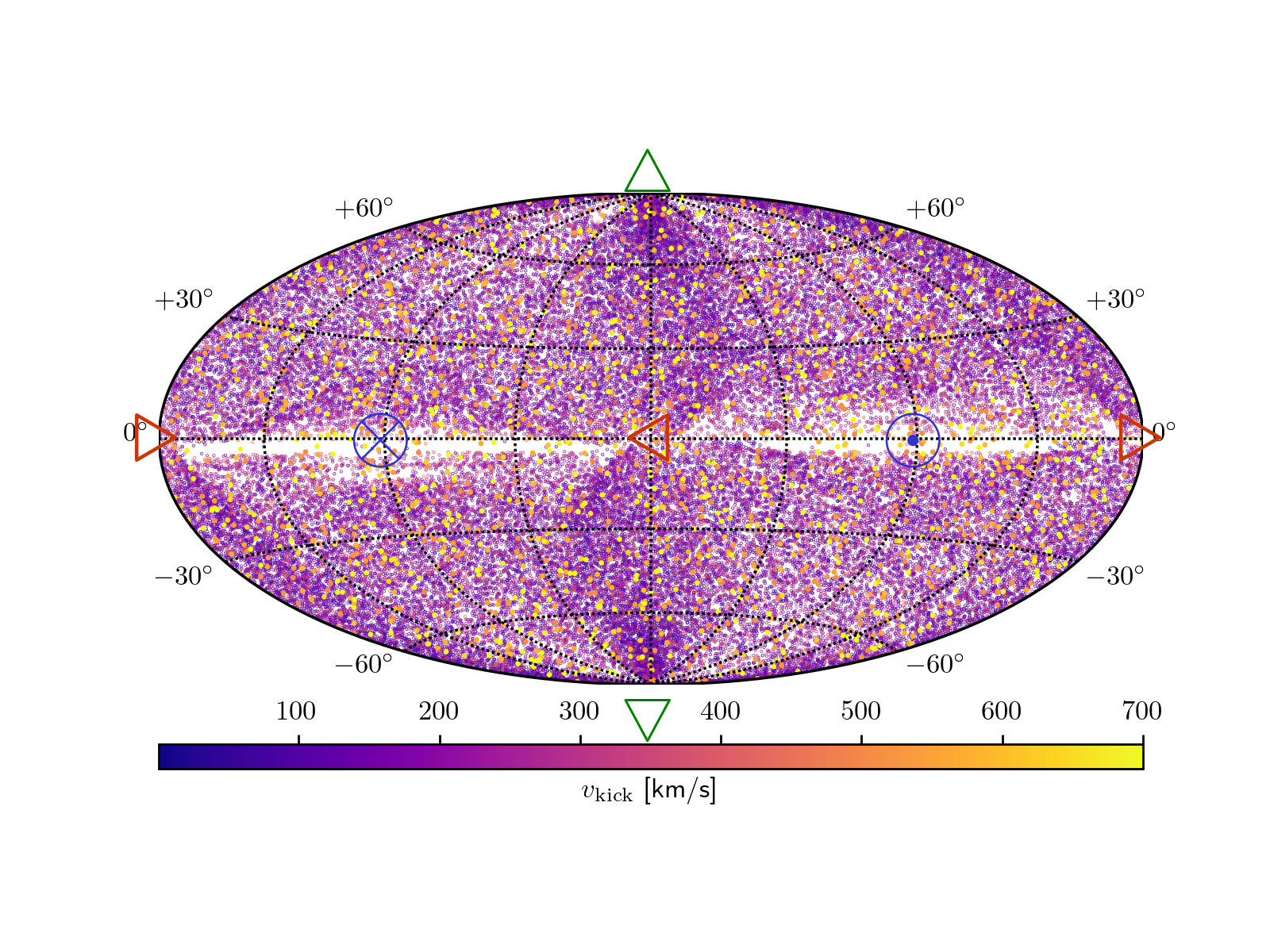}
		\caption{
			Top panel: distribution of breakup velocities of unstable triples for set L1 ($a_{\rm SMBH} = 0.1\pc$, dotted blue line), set L2 ($\abh = 0.01\pc$, dot-dashed green line) and L3 ($\abh = 0.01\pc$, dashed red line) and the corresponding isolated set (black histogram). Initial triples have $w=0$, $\aI = 1 \au$, $\aO = 2 \au$. The dotted blue, green and red vertical lines indicate the Hill velocity for $a_{\rm SMBH} = 0.1, 0.01$ and $0.005\pc$ (see Section~\ref{sec:discbound} for more details).
			Bottom three panels: Angular distribution of the breakup velocity vectors for the triples of sets L1, L2 and L3 in the reference frame co-rotating with the triple in its motion around the SMBH. The colour indicates the magnitude of the velocity in $\rm km/s$. Kicks with velocity greater than $350 \rm km/s$ are magnified. The green $\color{mygreen} \bigtriangleup $ and $\color{mygreen} \bigtriangledown$ indicate the radial and anti-radial directions, respectively. The red $\color{myred} \triangleleft$ and its antipode (red $\color{myred} \triangleright$) are the positive and negative normal to the orbit. The blue $\color{myblue} \odot$ and $ \color{myblue}\otimes$ correspond to the retrograde and prograde directions.
		}
		\label{fig:vkick_aI1_aO2}
	\end{figure}
	
	\begin{figure}[hbtp]
		\centering
		\includegraphics[width=0.9\linewidth]{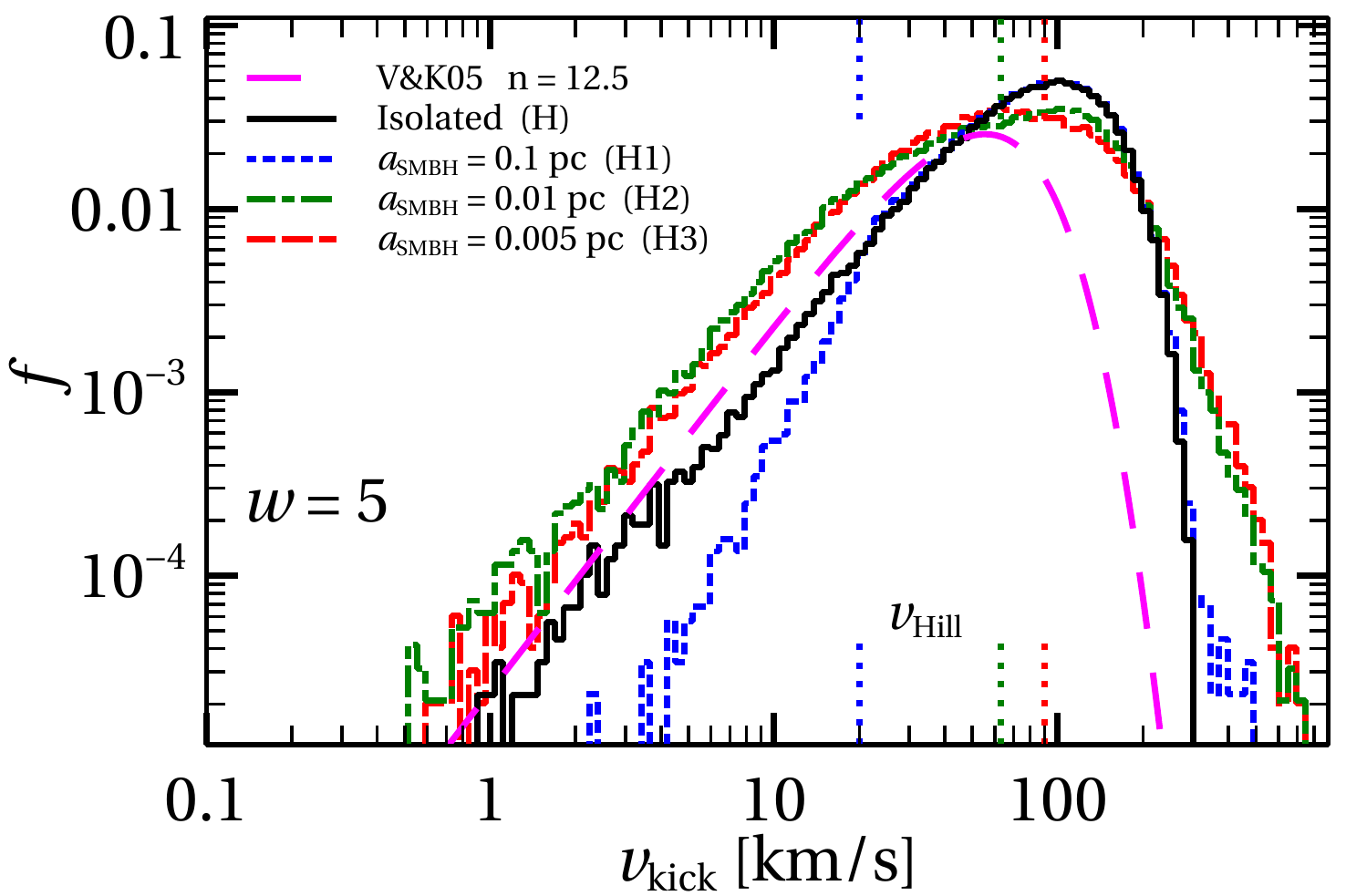}
		\includegraphics[width=0.9\linewidth,trim={1.6cm 1.9cm 1.3cm 1.9cm}, clip]{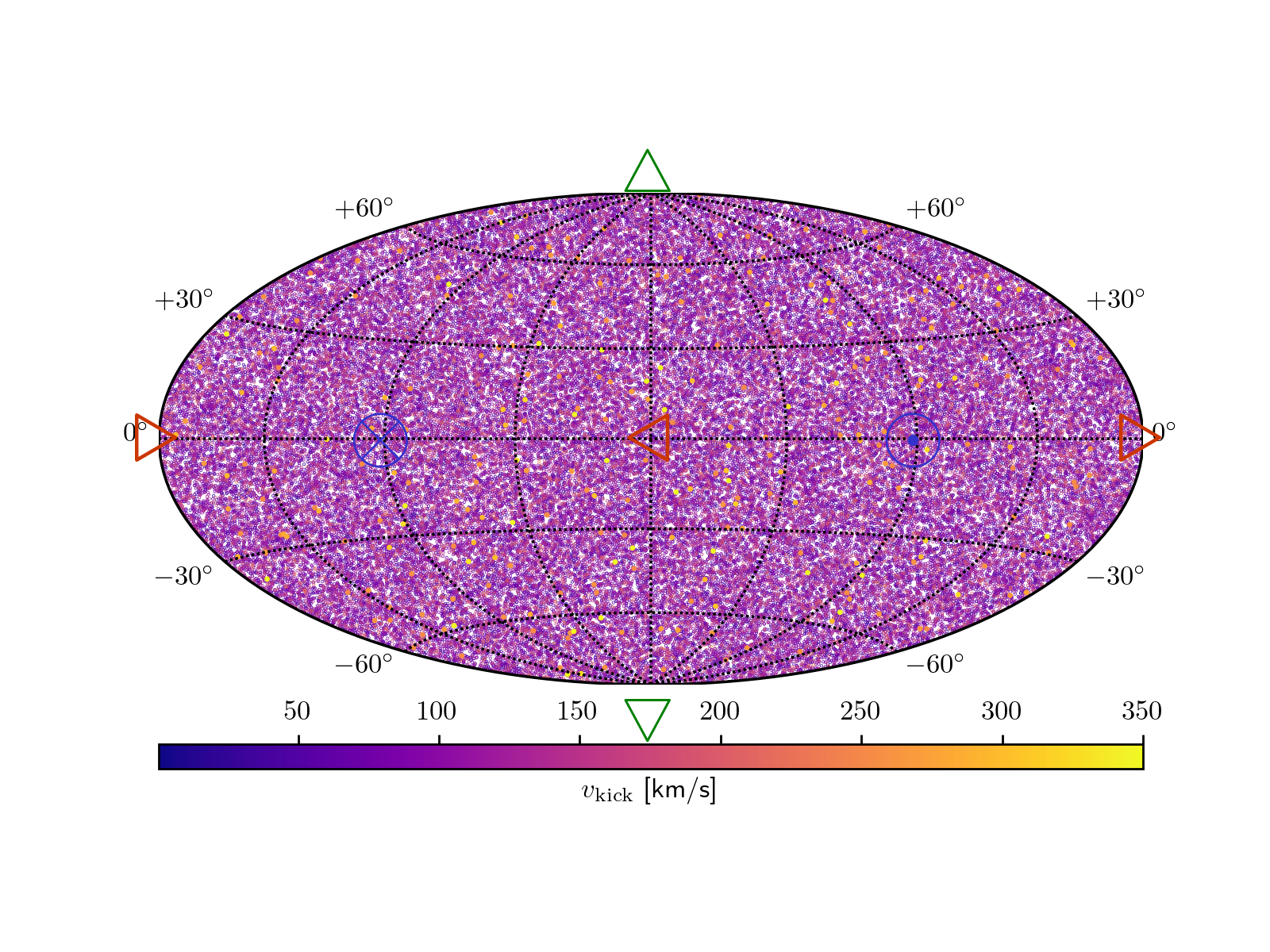}\\
		\includegraphics[width=0.9\linewidth,trim={1.6cm 1.9cm 1.3cm 1.9cm}, clip]{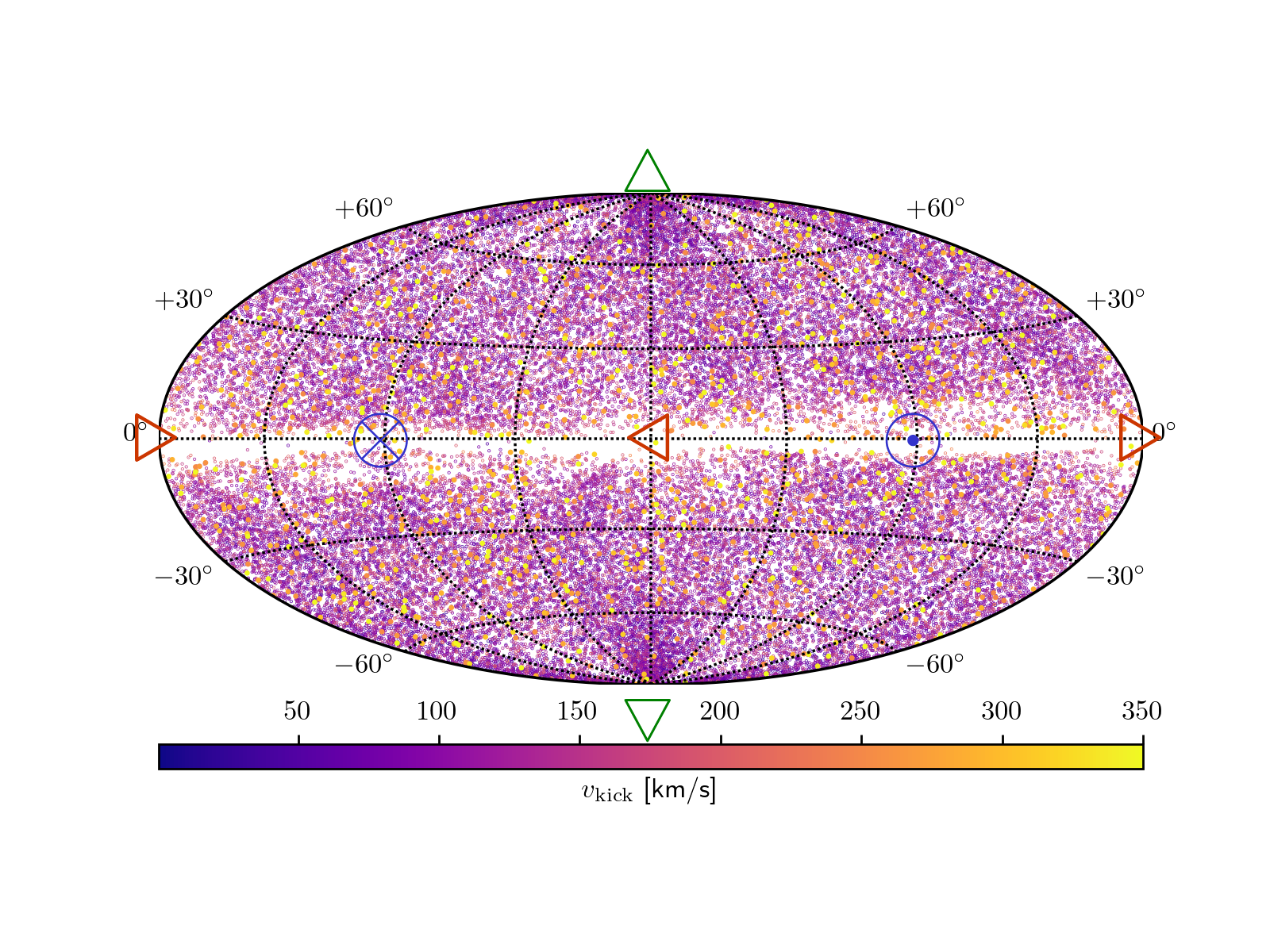}\\
		\includegraphics[width=0.9\linewidth,trim={1.6cm 1.9cm 1.3cm 1.9cm}, clip]{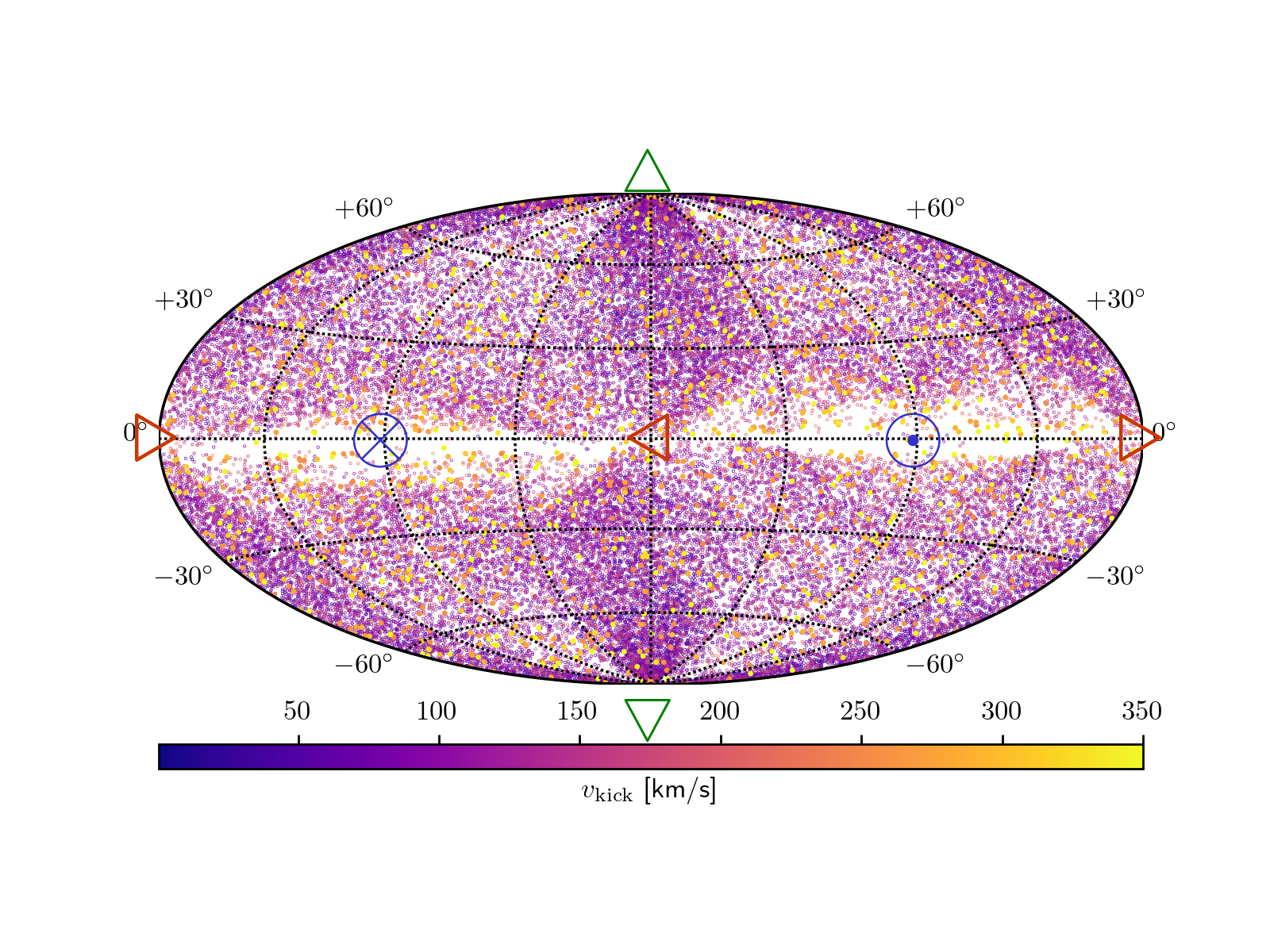}
		\caption{Same as Figure~\ref{fig:vkick_aI1_aO2}, but for set H1, H2 and H3 ($a_{\rm SMBH} = 1, 0.1$ and $0.01 \pc$) and the corresponding isolated set ($w=5$, $\aI = 1 \au$, $\aO = 2 \au$). In this case, kicks with velocity greater than $500 \rm km/s$ are magnified.
		}
		\label{fig:vkick_aI1_aO2_w5}
	\end{figure}

	\begin{table*}[ht]
		\begin{center}
			\caption{\footnotesize Outcomes of triple systems.\label{tab:outtri}}
			\begin{tabular}{lcccccccc}
				\hhline{=========}
				Properties \textbackslash{} Set &	L	 & L1 	  & L2     & L3      & 	H	  & H1		& H2	 & H3 	\\ \hline
				Collided (\%) 				& 10.443 & 8.802  & 2.520   & 0.908   &	0.110 & 0.097	& 0.127  & 0.133  	\\
				Not broken up (\%) 			&  0.185 & 0.317  & 0.001  & 0.002   &	1.160 & 2.561   & 0.109  & 0.030 	\\
				\hline
			\end{tabular}
		\end{center}
		
		\centering		\scriptsize 
		Row~1: percentage of collided systems;
		row~2: percentage of triples not broken up after 10000 dynamical times.
		
	\end{table*}
	
	Figure~\ref{fig:tbreak} shows the distribution of breakup times ($t_\mathrm{break}$) of the triple systems in isolation and in orbit around the SMBH. Overall, the tidal field decreases the lifetime of the triples, with the effect most apparent in the strong tidal field cases (sets L2 and H2 with $\abh = 0.01$, sets L3 and H3 with $\abh = 0.005 \pc$). The number of systems that have not yet broken up by the end of the integration is reported in Table~\ref{tab:outtri}. \rev{The mean number of orbits around the SMBH for a triple to break up is 0.05 in set L2 and 0.15 in set L3, while it is 0.29 and 0.53 and  in set H2 and H3, respectively.}
	
	The tidal field also affects the number of systems that result in a body-body collision between two particles of the triple (first row of Table~\ref{tab:outtri}). In triples with low angular momentum ($w=0$) the number of collided systems decreases steadily with increasing strength of the tidal field, down to ${\approx}1/10th$ of the number of equivalent collided systems in isolation (Table~\ref{tab:outtri}). The probability of a collision increases with the amount of time the system spends in a small volume of space. The decrease of number of collisions indicates that the system can break earlier without undergoing as many close encounters. This is also in agreement with the results of \citet{arca2018}, who found that the merger probability for black hole triplets is maximized for retrograde triples.
	
	In high angular momentum triples ($w = 5$), the number of collided systems is already very low even for triples in isolation, so the differences in number of collided systems is likely due to chaos, stochastic effects and small number statistics alone.

	\subsection{Magnitude and direction of breakup velocity kicks}
	Figures~\ref{fig:vkick_aI1_aO2} and \ref{fig:vkick_aI1_aO2_w5} compare the distributions of magnitude and direction of the breakup velocity kick for three-body systems in isolation and in circular orbits around the SMBH.
	
	Figure~\ref{fig:vkick_aI1_aO2} shows the results from triples with low total angular momentum.
	For $\abh = 0.1 \pc$, the breakup velocities are largely unaffected: the breakup direction is isotropic and the velocity distribution matches the one derived from the isolated case. On the other hand, deeper in the potential of the SMBH, for $\abh = 0.01$ and $0.005\pc$ the tidal field of the SMBH affects the breakup velocity both in magnitude and direction. 
	
	As the distance from the SMBH decreases, the breakup velocity decreases as well. Ejections occur preferably along the radial and anti-radial directions, while ejections along the prograde and retrograde directions are strongly disfavoured. Ejections towards the normal to the orbital plane are also disfavoured.
	
	Similar trends occurs for sets~H1, H2 and H3 (Figure~\ref{fig:vkick_aI1_aO2_w5}), whose triples have higher angular momentum compared to the L1, L2 and L3 sets. Compared to the low total angular momentum case, the breakup velocity is overall lower (top panel of Figure~\ref{fig:vkick_aI1_aO2_w5}) and the degree of anisotropy in the breakup direction is enhanced (bottom panels of Figure~\ref{fig:vkick_aI1_aO2_w5}). Additionally, a tail with velocities higher than the isolated case appears in sets H2 and H3, which was otherwise absent in sets L2 and L3.

\rev{	
	We also compare the breakup velocity distribution with that expected from statistical escape theory (equation 7.19 from \citealt{kart05}). Specifically,
}
	\begin{equation}\label{eq:vesc}
	f({v_\mathrm{kick}}) = \frac{(n-1)\left|E_\mathrm{tot}\right|^{n-1} m_\mathrm{sin}M_\mathrm{tot}/m_\mathrm{bin} v_\mathrm{kick} }{\left|E_\mathrm{tot}\right| + \frac{1}{2} ( m_\mathrm{sin}M_\mathrm{tot}/m_\mathrm{bin}) v_\mathrm{kick}^2}
	\end{equation}
\rev{where $m_\mathrm{sin}$ and $m_\mathrm{bin}$ are the masses of the single and the binary, respectively, and $n$ is a dimensionless parameter that depends on the total angular momentum of the system and that can be determined from numerical experiments. \citet{kart05} find that $n = 18\,L_\mathrm{norm}^2 + 3$, where $L_\mathrm{norm}$ is the total angular momentum of the system normalized to the maximum allowed one, corresponding to $w = 6.25$ \citep{mik94}. For $w=0$ we use $n = 3.5$, while for $w=5$ we set $n = 12.5$. Interestingly, the theoretical distribution fits well the distribution from isolated triples in the low-velocity regime, but fails to reproduce the high-velocity end for both L and H sets. In any case, the theoretical distribution does not fit at all the velocity kick distributions from triples in a tidal field.
}	
	
	\subsection{Orbital properties of the final binaries}
	
\begin{figure*}[hbt]
		\includegraphics[width=\linewidth]{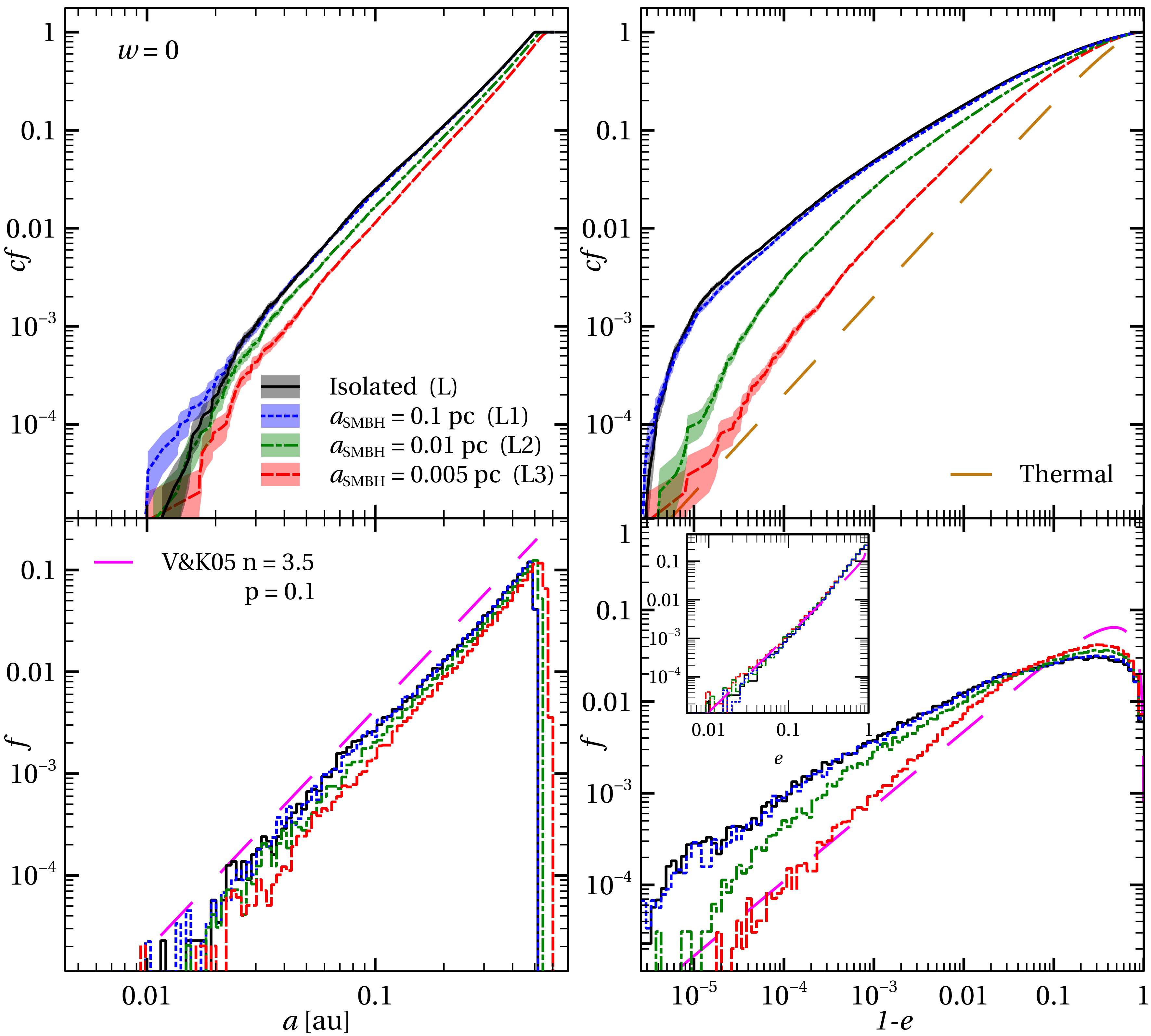}
		\caption{\footnotesize
			Top panels: cumulative distribution of semimajor axes (left) and eccentricities (right) of the binaries formed from the breakup of unstable triples in sets L1 (dotted blue line), L2 (dot-dashed green line), L3 (dashed red line), and the triples in isolation (solid black line). Yellow dashed line: thermal eccentricity distribution. Magenta dashed lines: Equations~\ref{eq:vkene} (left) and \ref{eq:vkene} (right), with $n = 3.5$ and $p = -0.2$, respectively.
			Bottom panel: same as top, but showing the corresponding non-cumulative distributions.}
		\label{fig:baecumL}
	\end{figure*}

\rev{
	Figure~\ref{fig:baecumL} shows the semimajor axes and eccentricity distributions of the binaries formed from the breakup of the triples with $w = 0$. 
	From statistical escape theory, we expect that the distributions of eccentricity and semimajor axis are given by (equations 7.26 and 7.31 from \citealt{kart05}): }
\begin{align}\label{eq:vkene}
& f(\left|E_\mathrm{b}\right|) = (n -1) \left|E_\mathrm{tot}\right|^{n-1} \, \left|E_\mathrm{b}\right|^{-n} \\
& f(e) = 2\,(p + 1)\, e \, (1 - e^2)^p \label{eq:vkecc}
\end{align}
\rev{	
	where $E_\mathrm{b} = -Gm_\mathrm{bin}/2a$ is the binding energy of the binary, and $p$ is a parameter, analogue to $n$, which can be computed from the empirical expression $p = L_\mathrm{norm}/2 - 1/4$ \citep{valt03}.
	
	The slope of the semimajor axis distribution of all sets is consistent with that \revi{expected from} statistical theory for all the L sets. 
	Here the most noticeable difference is that the semimajor axis distribution is increasingly shifted towards larger semimajor axes with decreasing distance from the SMBH. 
  }
	
	The eccentricity distribution presents more striking differences between weak and strong tidal fields. While in all cases the eccentricity distribution remains superthermal, it becomes increasingly thermal with decreasing distance from the SMBH (top right panel of Figure~\ref{fig:baecumL}).
	Note that the differences are mainly in the high-eccentricity tail: the distribution at low eccentricities does not present any differences and it is consistent with the expected theoretical distribution. On the other hand, the theoretical distribution does not match the high-eccentricity tail in any of the simulation sets, regardless of whether the triples are in isolation or not.
	
\begin{figure*}[hbt]
		\includegraphics[width=\linewidth]{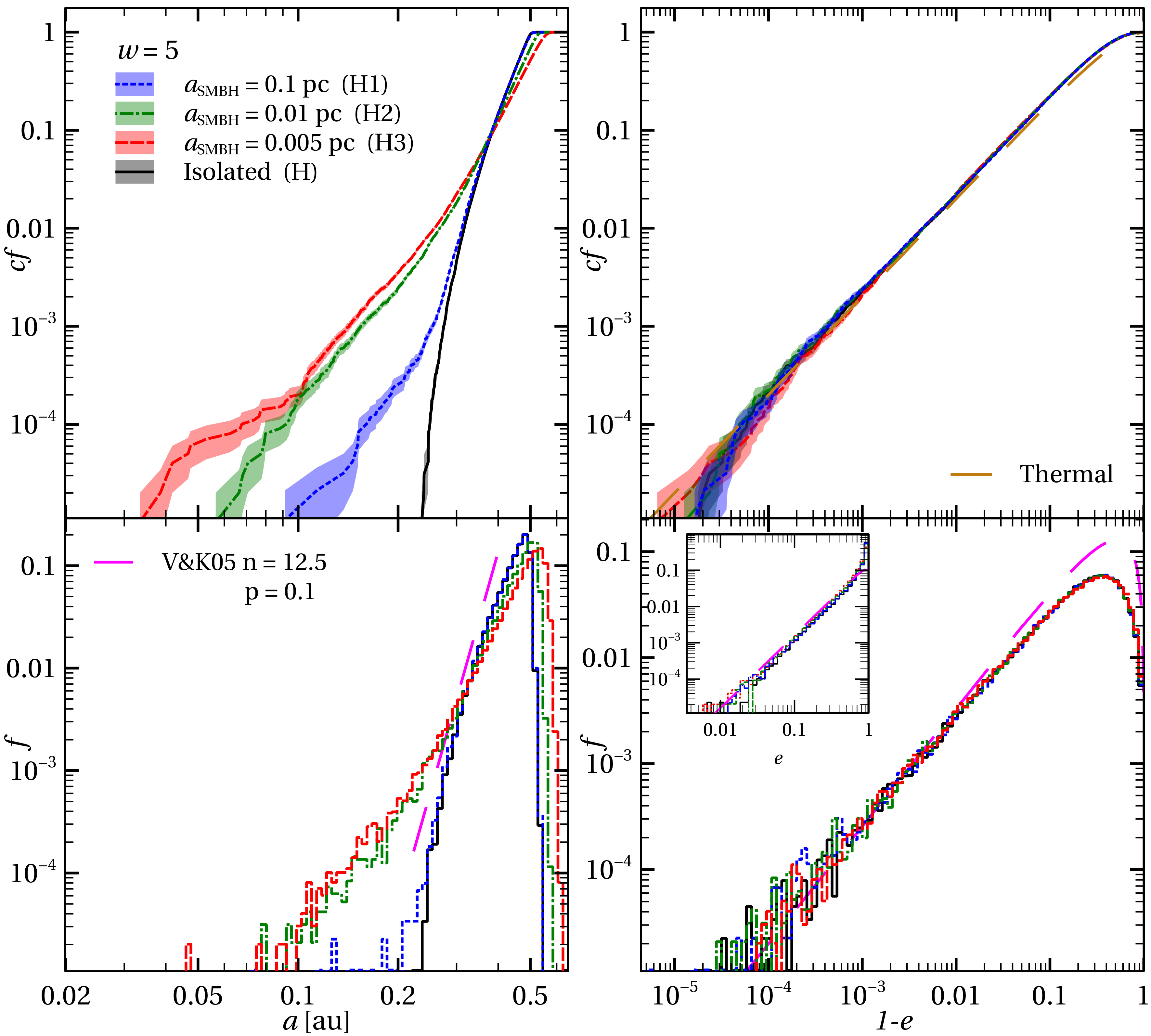}
		\caption{\footnotesize
			Top panels: cumulative distributions of semimajor axes (left) and eccentricities (right) of the binaries formed from the breakup of unstable triples in sets H1 (dotted blue line), H2 (dot-dashed green line), H3 (dashed red line), and the triples in isolation (solid black line). Yellow dashed line: thermal eccentricity distribution. Magenta dashed lines: Equations~\ref{eq:vkene} (left) and \ref{eq:vkene} (right), with $n = 12.5$ and $p = 0.1$, respectively. Bottom panel: same as top, but showing the corresponding non-cumulative distributions.}
		\label{fig:baecumH}
	\end{figure*}

	The situation is very different for binaries formed from the breakup of triples with high angular momentum ($w = 5$, H sets, Figure~\ref{fig:baecumH}).
	In this case, the eccentricity distribution is largely unaffected by the tidal field, and it is overall consistent with a thermal distribution, as also expected from statistical theory. In contrast, the semimajor axis distribution is strongly affected by the tidal field, deviating from the expected theoretical distribution. As shown in the bottom left panel of Figure~\ref{fig:baecumH}, the semimajor axis distribution extends towards smaller semimajor axes with decreasing distance from the SMBH. Note that the semimajor axis is on average larger than in the low angular momentum case. In this case, the tidal field enables the binaries to reach the small semimajor axis range allowed in low angular momentum triples, but that was otherwise inaccessible in high angular momentum triples. 
	
	As for the sets with low angular momentum, sets H2 and H3 also achieve larger semimajor axes with respect to the sets in isolation and in a weak tidal field (set H1). \rev{ Nonetheless, the binary apocenter remains inside the Hill radius, even for the H3 sets ($2\,a^\mathrm{max}_\mathrm{bin} \simeq 1.3 \au < 8.6 \au \simeq \Rh $).}
	
	\subsection{The role of the tidal field in the evolution of triples}\label{sec:discbound}
	
	Two common features emerge from the introduction of the Keplerian tidal field to an unstable three-body system, regardless of its initial angular momentum: the typical ejection velocity is lower, and the upper limit on the semimajor axis of the final binary increases. These two features are coupled, since the total energy of the triple is approximately conserved even in the presence of the tidal field. The total energy of the system after the breakup can be thus broken down into the sum of the (positive) energy of the escaper and the binding energy of the binary, or $\left|E_\mathrm{tot}\right| = E_\mathrm{esc} + E_\mathrm{bin}$. Consequently, the harder the binary, the higher the breakup velocity and \textit{viceversa}.
	
	The three-body evolution can be thought as a succession of episodes featuring strong interactions, in which all bodies interact ``democratically".  Most of these produce temporary ejections, in which one of the bodies gets ejected on a wide, bound orbit and the system can be regarded as an unstable, hierarchical triple.
	The three-body system goes through a succession of these states, until a single body acquires enough velocity to escape.
	
	The tidal field lowers the threshold velocity necessary for one body to escape the system. Therefore, in isolation an ejected body on a bound orbit will always return back to the binary, while in our case the tidal field might breakup the system before this occurs. 
	
	We can estimate it by considering the Hill radius as the maximum semimajor axis an ejected body can have with respect to the binary, beyond which the tidal forces from the SMBH will break the pair and the single body will not return to begin another strong interaction. The largest allowed semimajor axis of the final binary is easily derived:
\begin{equation}\label{eq:amax}
	a^\mathrm{max}_\mathrm{bin} = \left(\frac{(2\left|E_\mathrm{tot}\right|)}{G\,m_1 m_2} - \frac{1}{r_\mathrm{Hill}}\frac{m_1 m_2}{(m_1 + m_2)\,m_3}\right)^{-1}
\end{equation}
	where $m_1$ and $m_2$ are the masses of the binary components, and $m_3$ is the mass of the escaper. 
	
	For $r_\mathrm{Hill}\longrightarrow\infty$, Equation~\ref{eq:amax} reduces to $a^\mathrm{max}_\mathrm{bin} = {G\,m_1 m_2}/{2\left|E_\mathrm{tot}\right|}$, for which we obtain $a^\mathrm{max}_\mathrm{bin} \simeq 0.510 \au$, consistent with the value of $0.506 \au$ we obtain from the isolated set~L. For sets~L1,~L2~and~L3 Equation~\ref{eq:amax} leads to $a^\mathrm{max}_\mathrm{bin} \simeq 0.513, 0.538$ and $0.570\au$, respectively, which is in reasonable agreement with the results of the simulations ($0.508, 0.544$ and $0.588 \au$), given that the Hill region is not actually spherical (as shown by Figures~\ref{fig:vkick_aI1_aO2} and \ref{fig:vkick_aI1_aO2_w5}) and it does not constitute a "hard" boundary.
	
	In the high angular momentum simulation sets ($w = 5$, sets~H1,~H2~and~H3), $a^\mathrm{max}_\mathrm{bin}$ is overall slightly larger than in the low angular momentum sets; nonetheless Figure~\ref{fig:baecumL} presents the same qualitative trend described by Equation~\ref{eq:amax}.
	
	For the same reason, as $r_\mathrm{Hill}$ decreases, the breakup velocity becomes skewed towards lower velocities. This effect can be quantified by considering the Hill velocity of the triple system, defined as 
	\begin{equation}\label{eq:hillv}
	v_\mathrm{Hill} = \sqrt{G M_\mathrm{tot}/ r_\mathrm{Hill}}
	\end{equation}
	\rev{This velocity is the typical velocity of a loosely-bound test particle which orbits at the border of the Hill region.}
	In our setup, $v_\mathrm{Hill} \simeq 20, 63$ and $90 \kms$, for $\abh = 0.1 \pc$ (sets L1 and H1), $\abh = 0.01 \pc$ (sets H2 and L2) and $\abh = 0.005 \pc$ (sets L3 and H3), respectively. 
	
	The breakup velocity can be affected by the mechanism discussed above for $v_\mathrm{kick}<v_\mathrm{Hill}$. In this range, the distribution of breakup velocities shifts towards  lower velocities than $v_\mathrm{Hill}$, as shown by the top panel of Figure~\ref{fig:vkick_aI1_aO2} and \ref{fig:vkick_aI1_aO2_w5}.
	This also explains why the breakup velocity and especially its direction is more affected in the triples with high angular momentum ($w=5$), for which the breakup velocity is overall lower compared to the $w=0$.
	
	The angular distribution of the breakup velocity becomes anisotropic, following the shape of the Hill region around the triple. The triple can break more easily towards the radial and antiradial directions where the Hill region opens up. On the other hand, breakups are strongly disfavoured in the prograde and retrograde directions, where the energy threshold for the escape is larger. This anisotropy manifests itself also in the tidal breakups of high-mass ratio binaries in Keplerian potential \citep{tra16b}.
	
	Notice that in both Figures~\ref{fig:vkick_aI1_aO2}~and~\ref{fig:vkick_aI1_aO2_w5}, the distribution of $v_\mathrm{kick}$ for $\abh = 0.1 \pc$ is actually skewed towards higher velocities, compared to the isolated triples, which would imply a smaller value of $a^\mathrm{max}_\mathrm{bin}$. However, this is clearly not the case. We attribute the mismatch in the $v_\mathrm{kick}$ distribution to the impulse approximation that we assume to compute $v_\mathrm{kick}$ in the Keplerian case, which inevitably breaks down at low velocities. This also implies that the actual $v_\mathrm{kick}$ distribution is even more skewed at low velocities compared to the computed one shown in Figure~\ref{fig:vkick_aI1_aO2}~and~\ref{fig:vkick_aI1_aO2_w5}.

	The effects of the tidal field we have discussed so far do not depend on the initial angular momentum of the triple. 
	However, two angular momentum dependent features clearly emerge from Figure~\ref{fig:baecumL}~and~\ref{fig:baecumH}. First, the eccentricity distribution of the final binaries is affected by the tidal field only in the low angular momentum case. Second, binaries become tighter in the presence of the tidal field only in the high angular momentum case.
	
	Both effects are a consequence of the angular momentum transfer between the triple and its orbit around the SMBH. The superthermal eccentricity distribution in isolated, low angular momentum triples results from the fact that, in order for a breakup to occur, the three particles need to end up in a sufficiently small volume to accelerate one body to ejection. This translates to lower angular momentum interactions tending toward being more compact, more eccentric and having higher ejection velocities, relative to higher angular momentum interactions, and results in the eccentricity distribution of the final binaries being superthermal.
	
\begin{figure}[hbt]
	\includegraphics[width=1\linewidth]{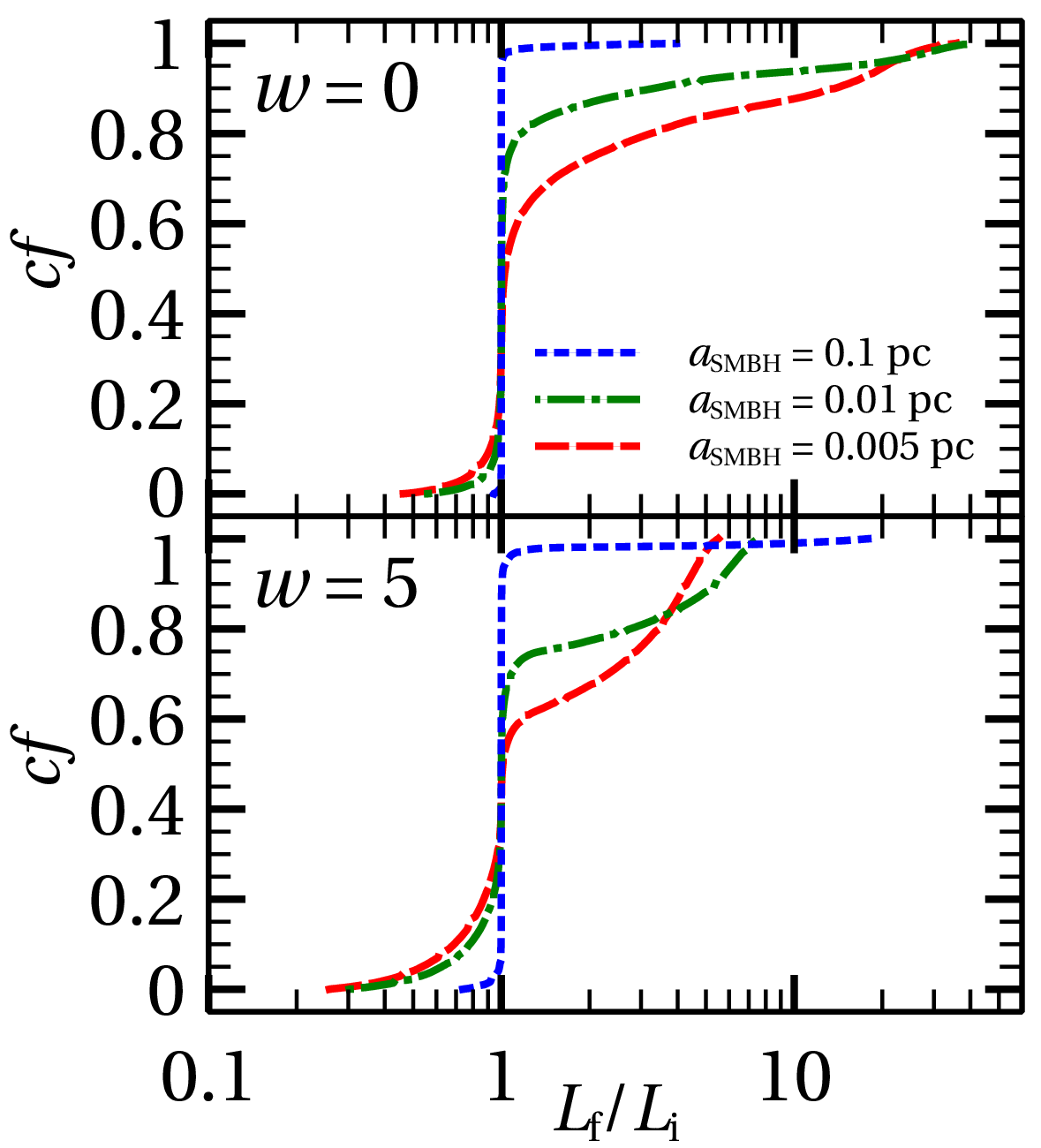}
	\caption{\footnotesize
		Cumulative fraction of the ratio between angular momentum of the triple before breakup ($L_\mathrm{f}$) and initial angular momentum of the triple ($L_\mathrm{i}$). Dotted blue line: $a_{\rm SMBH} = 0.1\pc$; Dot-dashed green line: $\abh = 0.01\pc$; dashed red line: $\abh = 0.01\pc$. Top panels: sets L1, L2 and L3 ($w = 0$). Bottom panel: sets H1, H2 and H3 ($w = 5$). Sets L and H are not shown, since there is no change in the total angular momentum for triples in isolation.
	}
	\label{fig:deltal}
	\end{figure}

	The tidal field torques the system, transferring angular momentum from the orbit around the SMBH to the triple system, as shown in Figure~\ref{fig:deltal}. Low angular momentum triples mostly gain angular momentum, which results in the final eccentricity distribution approaching the thermal distribution. The angular momentum gain does not affect high angular momentum triples, since those systems begin with enough angular momentum to be redistributed among the particles, so that the final eccentricity distribution is always thermal.
	
\begin{figure}[hbt]
	\includegraphics[width=1\linewidth]{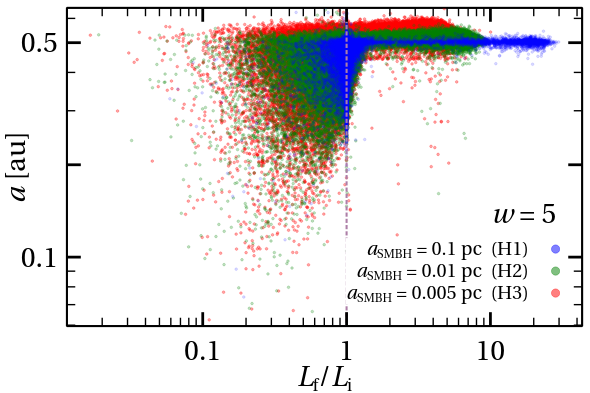}
	\caption{\footnotesize
		\rev{
		Final semimajor axis of binaries produced from triple breakup versus the ratio between the angular momentum of the triple before breakup  ($L_\mathrm{f}$) and the initial angular momentum of the triple  ($L_\mathrm{i}$). Only the sets with $w=5$ are shown. Blue circles: $a_{\rm SMBH} = 0.1\pc$ (H1); green circles: $\abh = 0.01\pc$ (H2); red circles: $\abh = 0.01\pc$ (H3). The dotted vertical line indicates the region of no angular momentum exchanges.}
	}
	\label{fig:adeltal}
	\end{figure}

	However, the transfer of angular momentum can also go the other way around: from the triple to the orbit around the SMBH. This effect is obviously more important for high angular momentum triples, and actually makes them behave more like low angular momentum triples. This is clear from Figure~\ref{fig:adeltal}, which shows that the binaries with smaller semimajor axes \revi{are formed from the triples that have lost most angular momentum}.
		
\section{Binary-single encounters in a Keplerian potential}\label{sec:binsin}

\subsection{Loose binaries: $a_\mathrm{bin} \simeq r_\mathrm{Hill}$}

	\begin{table}[hb]
	\begin{center}
		\caption{\footnotesize Outcomes of binary-single encounters.\label{tab:outbin}}
		\begin{tabular}{lccc}
			\hhline{====}
			Set \textbackslash{} Outcome &	Ionization (\%)	 & Flyby (\%) 	  & Exchange (\%) \\ \hline
			E1is  						 &    1.766 &  96.994  &  1.24 \\
			E1  						 &    41.514 &  58.135  &  0.398  \\
			E1p  						 &    85.078 &  14.563  &  0.388  \\\vspace{2pt}
			E1r						 &    8.853 &  90.826  &  0.342  \\
			
			E2is  						 &  5.662 & 90.216  & 4.122  \\
			E2  					     &  14.146 & 86.422  & 0.172  \\
			E2p  						 &    12.981 &  86.941  &  0.079  \\
			E2r						 &    12.989 &  86.925  &  0.087  \\
			\hline
		\end{tabular}
	\end{center}
\end{table}

\begin{figure*}[hbt]
	\includegraphics[width=\linewidth]{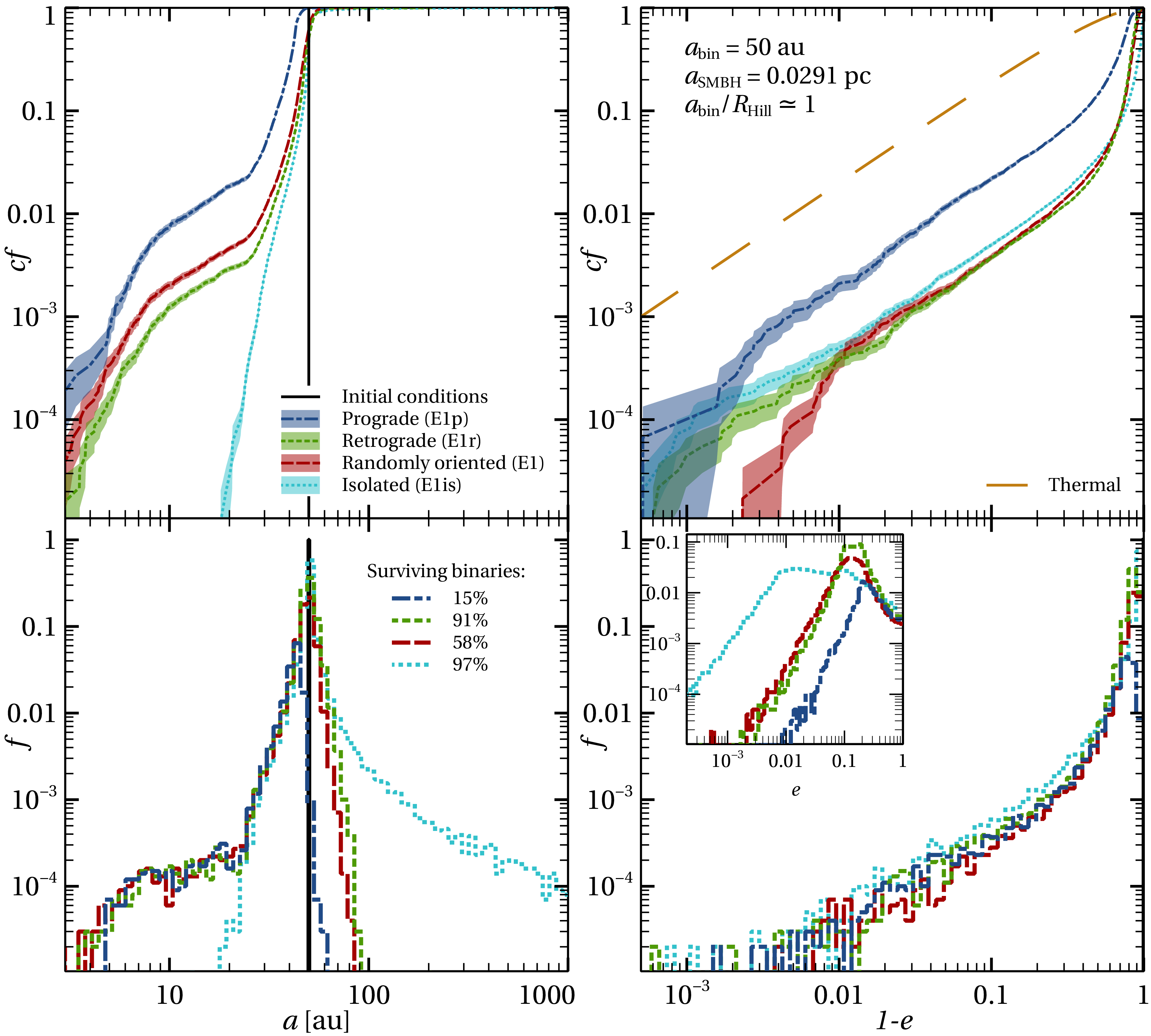}
	\caption{\footnotesize
		Top panels: cumulative distributions of semimajor axes (left) and eccentricities (right) of the binaries that survive three-body encounters for sets E1 (dashed red line), E1p (dot-dashed blue line), E1r (dotted green line), the corresponding set in isolation E1is (fine dotted cyan line) and the initial conditions (solid black line). Yellow dashed line: thermal eccentricity distribution.
		Bottom panel: same as top, but showing the corresponding non-cumulative distributions. The initial eccentricity is zero so it does not appear in the plot.}
	\label{fig:aecumE1}
\end{figure*}

\begin{figure*}[hbt]
	\includegraphics[width=\linewidth]{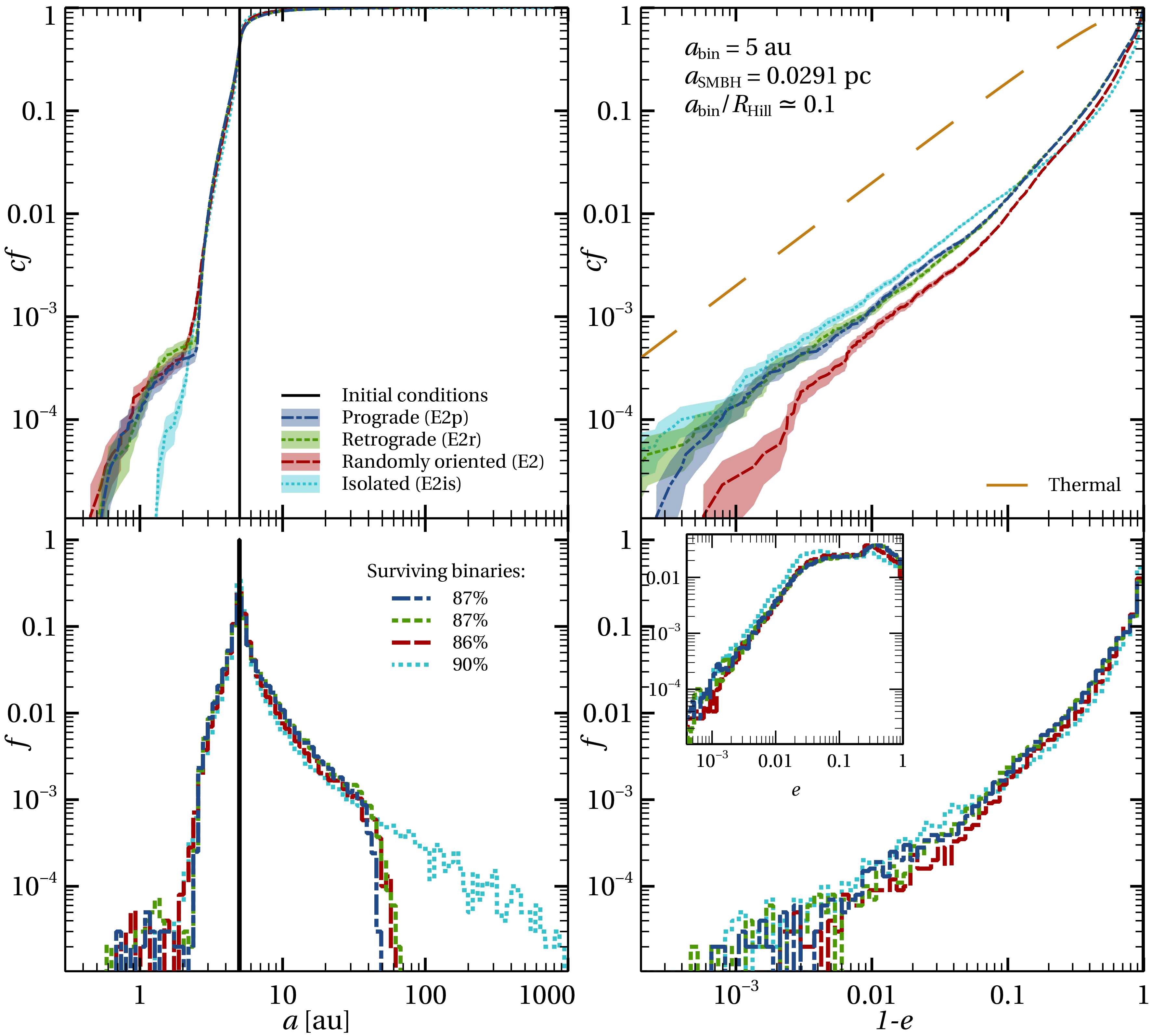}
	\caption{\footnotesize
		Top panels: cumulative distributions of semimajor axes (left) and eccentricities (right) of the binaries that survive three-body encounters for sets E2 (dashed red line), E2p (dot-dashed blue line), E2r (dotted green line), the corresponding set in isolation E2is (fine dotted cyan line) and the initial conditions (solid black line). Yellow dashed line: thermal eccentricity distribution.
		Bottom panel: same as top, but showing the corresponding non-cumulative distributions. The initial eccentricity is zero so it does not appear in the plot.}
	\label{fig:aecumE2}
\end{figure*}

\rev{
	Figure~\ref{fig:aecumE1} shows eccentricity and semimajor axis distributions of the binaries after the three-body encounters in sets~E1 (isotropic binary orientations), E1p (prograde binaries), E1r (retrograde binaries) and Eis (isolated encounters). In these sets, the binary is initially circular with a semimajor axis of $\abin = 50 \au$, which approximately corresponds to the Hill radius of the binary system at this distance from the SMBH (see Equation~\ref{eq:hill}).
	
	The tidal field has a strong impact on the fraction of surviving binaries, as well as on their orbital properties. The outcome of the encounter also strongly depends on the initial inclination of the binaries. 
	
	Table~\ref{tab:outbin} shows the outcome fraction of the binary-single encounters.
	About $85\%$ of the binaries from set E1p end up ionized by the encounter, while only $9\%$ gets ionized in set E1r. This is not surprising, since it is very well known that retrograde binaries are more stable with respect to prograde ones, and can achieve larger semimajor axes without getting disrupted by the tidal field \citep{hen70,inn80,ham91,inn97,geo13,tra16,gri17}. 
	This emerges clear from the semimajor axis distribution, which gets truncated at about $60\au$ in set E1p, in contrast with the upper limit of $90\au$ in set E1r. In set E1is, since there is no tidal field, the binary semimajor axis can be arbitrarily large and only about $3\%$ of the binaries get ionized by the tidal field. 

	Despite the disruptive effect of the tidal field, the binaries in the Keplerian sets become more hard compared to the binaries in isolation. 
	The semimajor axis distribution reaches significantly lower values in sets~E1, E1p and E1r with respect to the isolated case, indicating that Keplerian three-body encounters are more efficient at hardening the binaries. The number of binaries with smaller semimajor axes than in isolation remains nonetheless small, about a few \% of the total number of survived binaries.
	
	The final eccentricity distribution has approximately the same range independently of whether the encounter occurs in isolation or around the SMBH. However, the eccentricity distribution of set E1p is significantly more skewed towards higher eccentricities compared to those of the isolated set, with the median eccentricity being almost 10 times higher than the in the isolated case ($\widetilde{e}_\mathrm{E1p}\simeq 0.26$ vs $\widetilde{e}_\mathrm{E1is}\simeq0.03$). Also binaries from sets E1r and E1 are on average more eccentric than in isolation with a median of  $\widetilde{e}_\mathrm{E1}\sim \widetilde{e}_\mathrm{E1r}\simeq0.14$ compared to $\widetilde{e}_\mathrm{E1is}\simeq0.03$.
}
\subsection{Tight binaries: $a_\mathrm{bin} = 0.1 r_\mathrm{Hill}$}	

\rev{
	Figure~\ref{fig:aecumE2} shows the same as Figure~\ref{fig:aecumE1}, but for the E2~sets. In this case, the binary has an initial semimajor axis of $a_{\rm bin} = 5 \au$, much lower than the Hill radius of the binary. 
	
 	In these sets, the binary orientation does not influence the surviving binary fraction, which is close to the isolated case. The tidal field does not participate in the breakup of the binaries, but still affects their final orbital parameters.
	
	The semimajor axis distribution is nearly the same for all Keplerian sets. As in the E1~sets, the minimum semimajor axis in the Keplerian sets is smaller than in the isolated case, although only a few $0.1\%$ of the binaries reach a semimajor axis smaller than the lower minimum of $0.6\au$ of set~E1is.

	The eccentricity distribution is the same for sets E2p and E2r, while binaries from set E2 are overall less eccentric. The median eccentricity of the E2is set is about $\widetilde{e}_\mathrm{E2is}\simeq 0.1$, compared to $\widetilde{e}_\mathrm{E2p}\sim \widetilde{e}_\mathrm{E2r}\simeq 0.18$ of the sets~E2p and E2r, and $\widetilde{e}_\mathrm{E2}\simeq0.15$ of set E2. Overall, we find more binaries in sets E2, E2p and E2r with moderate eccentricities ($e=0$--$0.9$) than in set E2is.}
	

\subsection{The role of the tidal field \revi{during} binary-single encounters}	
	
\rev{
	The Keplerian potential generated by the SMBH introduces two more forces with respect to isolated encounters: the tidal force and the Coriolis force. The first can be expressed in the reference frame rotating with the binary mean motion around the SMBH as 
	\begin{equation}\label{eq:ftid}
	\mathbf{F}_\mathrm{tid} = \Omega_\mathrm{SMBH}^2 (3\mathbf{x} - \mathbf{z})
	\end{equation}
	where $\Omega_\mathrm{SMBH}$ is the angular velocity around the SMBH and $\mathbf{x}$ and $\mathbf{z}$ are the rotating coordinates, in which $\mathbf{x}$ extends along the radial direction and positive $\mathbf{z}$ is normal to the plane of the orbit. Therefore, the tidal force is mostly independent of the orientation of the binary with respect to its orbit around the SMBH. The Coriolis force can be instead expressed as 
	\begin{equation}
	\mathbf{F}_\mathrm{Cor} = -2\mathbf{\Omega}_\mathrm{SMBH}\times\mathbf{v}_\mathrm{bin}
	\end{equation}
	where $\mathbf{v}_\mathrm{bin}$ is the velocity vector of a star in the rotating reference frame, and strictly depends on the orientation of the binary. Specifically, the Coriolis force will be centripetal for retrograde binaries and centrifugal for prograde ones.
	
	Since $F_\mathrm{Cor} \propto v_\mathrm{bin} \propto a_\mathrm{bin}^{-1/2}$, the Coriolis force is more effective in tight binaries. Moreover, since the Coriolis force cannot do work, it will not change the energy of the system. However, by changing its angular momentum, it can still affect the overall dynamics, and thus alter the final energy state of the system.
	In contrast, for the tidal force $F_\mathrm{tid} \propto a_\mathrm{bin}$, such that it is stronger in loose binaries, and can directly affect both energy and angular momentum.
			
	For these reasons, a greater discrepancy arises in the final semimajor axis distribution for the loose binaries of set~E1 when comparing Keplerian and isolated encounters, while little differences are present in tight binaries of set~E2 that reside deep within their Hill sphere. 
	
	Eccentricity, on the other hand, is affected regardless of the $a_\mathrm{bin}/R_\mathrm{Hill}$ ratio, and it is more sensitive to the initial orientation of the binary. The Coriolis force determines the stability and thus the survival of the loose binaries of set E1. On the other hand, the tight binaries of set E2 reside deep in their Hill region so that they are stable independently of their orientation. 
	
	Another interesting result is due to this fact: the binaries from sets E2p and E2r have the same eccentricity distribution after the encounter, and are on average twice as eccentric than the isotropically oriented binaries of set E2. In other words, non-coplanar binaries become much less eccentric than coplanar ones. This might seem counter-intuitive at first, since the Kozai-Lidov mechanism, which can bring binaries to extremely high eccentricities, is efficient at high inclinations. However, with this setup we are looking at the immediate outcome after the encounter, and not at the long-term secular dynamics of the binary around the SMBH.
}
	
	
	
	\section{Implications for gravitational wave mergers}\label{sec:gw}
	
	\subsection{Unstable triples (hard binaries)}
	
	\begin{figure}[hbt]
		\includegraphics[width=\linewidth]{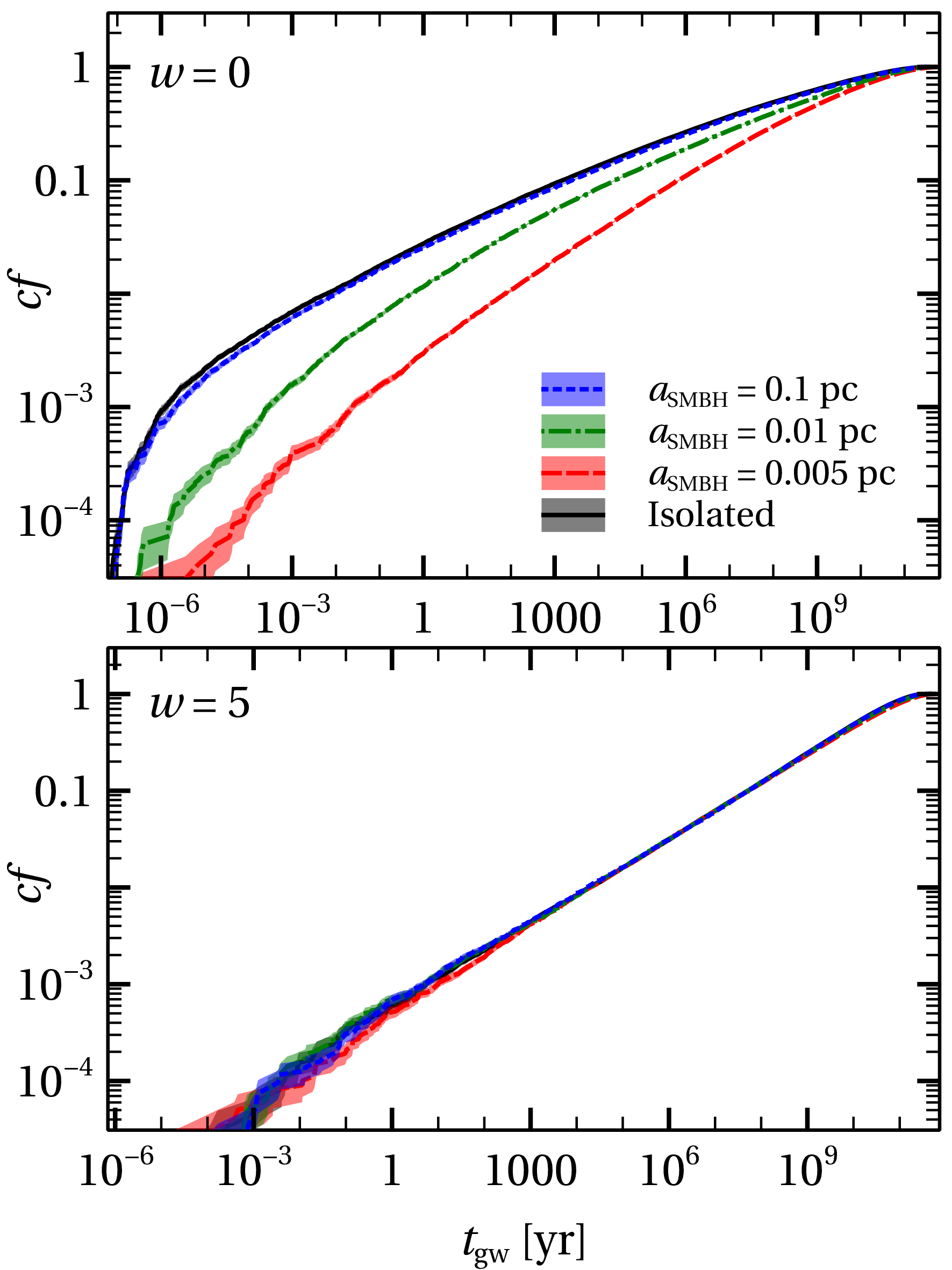}
		\caption{\footnotesize
			Cumulative distribution of gravitational wave coalescence timescale for surviving binaries in all sets of simulations of unstable hierarchical triples.
			Top panel: sets with $w = 0$ (L1, L2, L3 and L). Bottom panel: sets with $w = 5$ (H1, H2, H3 and H). Dashed red line: $\abh = $ oriented binaries; dotted blue line: prograde binaries; dot-dashed green line: retrograde binaries, fine dotted cyan line: sets in isolation; black line: initial coalescence timescale. Here we consider only binaries with merger timescale shorter than the initial one. The distribution are normalized to the number of binaries in the isolated sets.
		}
		\label{fig:gwt_bound}
	\end{figure}
	
\rev{
	Figure~\ref{fig:gwt_bound} shows the gravitational wave merger timescales calculated from the eccentricity and semimajor axis of the binaries formed from the simulations of unstable hierarchical triples \citep{peters64}. As explained in Section~\ref{sec:methods}, we can also interpret these simulations as the result of hard binary encounters. This type of encounter is typical of AGN disks, where the relative velocity between the binary and the single within the disk is small due to the coherent Keplerian motion \citep{secunda18,lei16a,lei18,mcker18,tagawa19}.
	
	In all sets, the merger timescale is dominated by the final eccentricity of the binary. Isolated, low angular momentum ($w=0$) triples have the shortest merger times due to the superthermal eccentricity of the final binaries. The addition of the tidal field makes the triples gain angular momentum from the orbit, so that the eccentricity distribution becomes more thermal and the average merger time increases.
	
	Since the final eccentricity of high angular momentum ($w=5$) triples is always thermal, there is no difference in the merger time distribution, regardless of the presence of the Keplerian tidal field.
}

	\subsection{Binary-single encounters (soft binaries)}
\begin{figure}[hbt]
		\includegraphics[width=\linewidth]{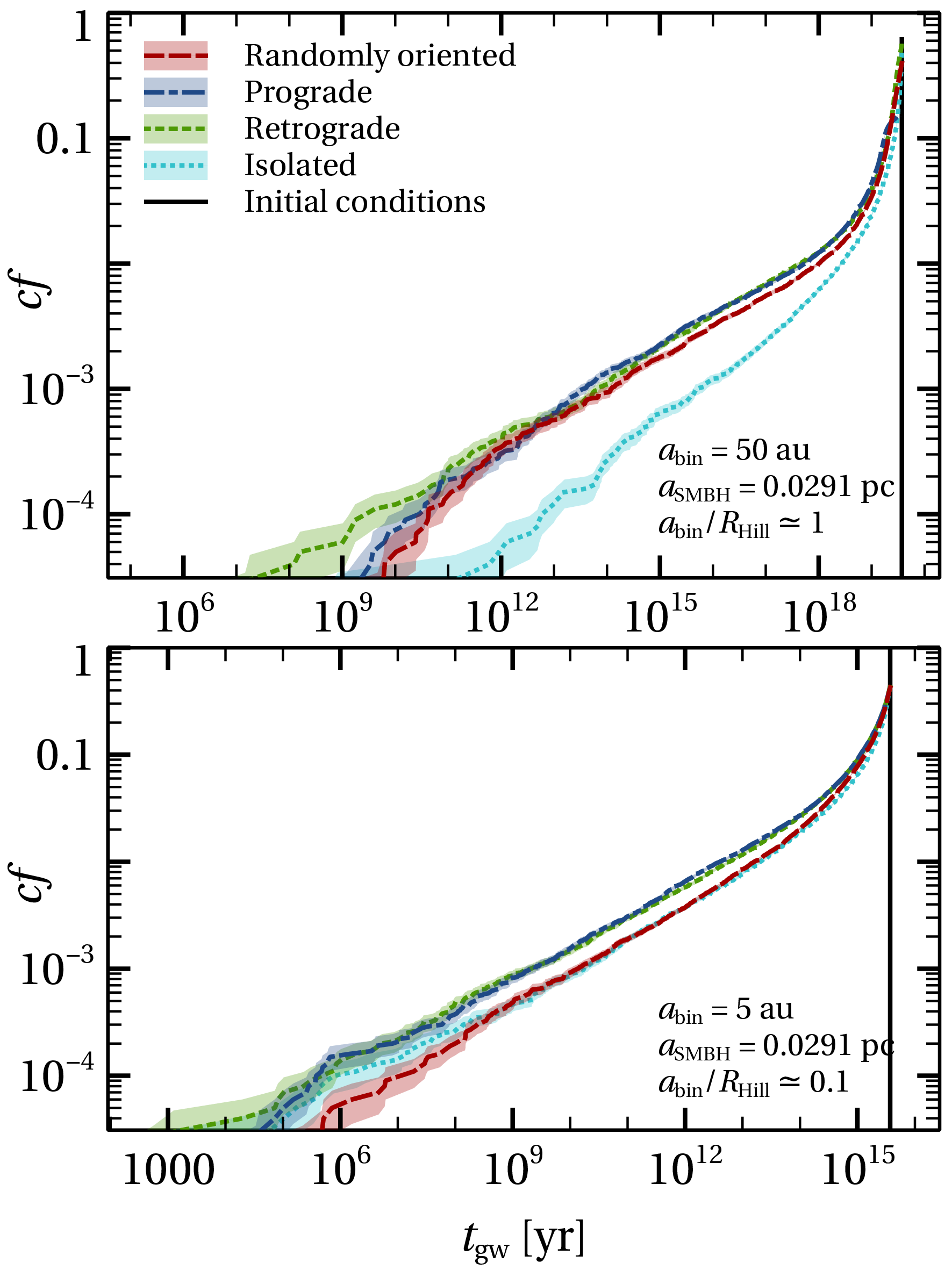}
		\caption{\footnotesize
			Cumulative distribution of gravitational wave coalescence timescale for surviving binaries in all sets of simulations of binary-single encounters.
			Top panel: sets with $a_{\rm bin} = 50 \au$ (E1, E1r, E1p and E1is). Bottom panel: sets with $a_{\rm bin} = 5 \au$ (E2, E2r, E2p and E12s).  Dashed red line: random oriented binaries; dot-dashed blue line: prograde binaries; dotted green line: retrograde binaries, fine dotted cyan line: sets in isolation; black line: initial coalescence timescale. Here we consider only binaries with merger timescale shorter than the initial one. The distribution are normalized to the number of binaries in the isolated sets.
		}
		\label{fig:gwt}
	\end{figure}

\rev{
	Figure~\ref{fig:gwt} shows the cumulative distribution of the coalescence timescale $t_\mathrm{gw}$ for the surviving binaries of binary-single encounters. In all the simulated sets, the binary is always soft with respect to the encountering body. This type of encounter typically occurs in high density, spherical cusps of compact remnants around SMBHs.
	
	For both wide and tight binaries, the encounter shortens the coalescence timescale by over $9$--$12$ orders of magnitude. This is a common feature of three-body encounters, such as those occurring in globular and young star clusters \citep[e.g.][]{zio14,tra14,doraz18,ant18b,sams18a,sams18b,sams18c,baner18,arc19b,sams19a,sams19b,kuma19,diocarlo19,rast19,rodkarl19c}.
	
	 This effect is further enhanced by the inclusion of the tidal field of the SMBH. The average decrease in coalescence timescale after the encounter is 26\% in set~E1 (8\% in set~E1is), and by 29\% in set~E2 (21\% in set~E2is). Including the tidal field is more important for loose binaries: in set~E1 and E2 there are $56\%$ and $21\%$ more binaries with $t_\mathrm{gw}$ less than half the initial one than in set E1is and E2is, respectively. The decrease in merger time is particularly enhanced in set E1p ($121\%$ more binaries with $t_\mathrm{gw}$ less than half the initial one) and sets E2p and E2r (about $36\%$ more binaries $t_\mathrm{gw}$ less than half the initial ones).

	Since the initial semimajor axis of the binaries remains quite high, only a few hundred binaries from set E2 merge within a Hubble time after the encounter. We find that the Keplerian tidal field can almost double the probability of mergers, with $85\%$ more binaries merging within $13.8 \, \mathrm{Gyr}$ in set~E2p than in set E2is. Note that since the smallest semimajor axis from set~E2 is about $1\au$, thus the decrease in merger time is mostly due to the high eccentricity.
}

\rev{
	We stress that our simulations do not include post-Newtonian corrections, whose effect is to shorten the merger timescale and increase the number of mergers in the timeframe of the simulations. This effect will likely be enhanced in the Keplerian scenarios, since the tidal field tends to excite eccentricities, leading to closer encounters. Therefore, our estimate is to be considered as a lower limit.
	
	Several authors have investigated mergers of compact binaries around SMBHs \citep[e.g.][]{lei18,arc18b,gond18,mcker18,arc19a,rass19,mcker19}, many focusing mostly on the role of the Kozai-Lidov mechanism \citep[e.g.][]{ant12,vanlan16,pet17,hoa18,ham18,zhan19,hoa19}. In particular, \citet{hoa18} identify two merger channels: one induced by the eccentric Kozai-Lidov mechanism, the other induced by gravitational wave radiation only. Our findings imply that three-body encounters can dramatically affect both merger channels. Encounters can shorten the gravitational merger timescale by exciting the eccentricity and shrinking the binary semimajor axis, but also trigger the Kozai-Lidov mechanism by altering the orientation of the binaries with respect to their orbital plane around the SMBH.
	
	In general, we expect that taking into account encounters will lead to an increased merger rate with respect to studies that take into consideration only Kozai-Lidov oscillations \citep[][]{lei18}. Furthermore, we predict that the number of interactions needed to harden a black hole binary until it can coalesce can be substantially lower than expected by not taking into account the tidal field of the SMBH. Quantifying the enhancement of merger rates due to encounters requires a thorough set of Monte Carlo simulations to adequately sample the relevant parameter space, which goes beyond the scope of the current paper.
}

	\section{Summary}\label{sec:conclusions}
	
	In this work, we investigate the impact of the Keplerian tidal field on the dynamics of three-bodies in the vicinity of a SMBH, across the spectrum of relative velocities: bound unstable triples (low velocity dispersion) and binary-single three-body scatterings (high velocity dispersion).
	
 	We run highly accurate four-body simulations of two different scenarios:
	a) a three-body encounter between a binary and a single body orbiting a SMBH and b) an unstable triple that decays into a binary and a single body while orbiting around a SMBH.
	We then re-run the same scenarios in isolation, and compare the outcomes.
	
	Our findings are summarized below.
	\begin{itemize}
		\item \textbf{Breakup of unstable triples}. Unstable triples are affected by the tidal field when the typical breakup velocity of the system in isolation is smaller than the corresponding Hill velocity (Equation~\ref{eq:hillv}). When this occurs, angular momentum is exchanged between the triple and the orbit around the SMBH, which changes the properties of the breakup. Low angular momentum triples gain angular momentum from the orbit and the eccentricity distribution of the final binaries becomes more thermal (from superthermal). Conversely, high angular momentum triples can lose angular momentum, which makes them behave more like low angular momentum triples such that the final semimajor axis becomes remarkably smaller. Regardless of the angular momentum exchange, the upper limit of the final semimajor axes increases, because the system becomes tidally limited (Equation~\ref{eq:amax}). \rev{We also report a mismatch of the final binary eccentricity and breakup kick distributions of isolated triples with those expected from statistical escape theory, which deserves future attention.}

		\item \textbf{Binary-single encounters}. The tidal field alters the orbital properties of the binaries after the encounter. Binaries that are particularly affected are those whose semimajor axis is close to the Hill radius of the system, and prograde binaries. The outcome depends on how deep the binary resides in its Hill sphere. Binaries close to their Hill radius tend to become more eccentric and to harden more than in isolation. This effect is present also in binaries that reside deep in their Hill radius, although to a lesser degree. Depending on the initial conditions, the inclusion of the tidal field can double the merger probability of binary black holes.
		
	\end{itemize}

	Finally, our results demonstrate that Keplerian three-body encounters can have an important role in triggering the coalescence of compact binaries around SMBHs: a single encounter can decrease the gravitational wave merger timescale by orders of magnitude. \rev{The inclusion of the Keplerian tidal field can almost double the merger probability compared to the isolated case}. More quantitative predictions on the production of gravitational wave sources around SMBHs require a dedicated population synthesis study that will be presented in the next work of the series.
	
	\acknowledgements
	It is a pleasure to thank Alessandro Peloni for helpful discussions on the MOID. This work was supported by JSPS KAKENHI Grant Numbers 17F17764, 19H01933 and 17929016. MS acknowledges funding from the European Union’s Horizon 2020 research and innovation programme under the Marie-Sklodowska-Curie grant agreement Number 794393.  NWCL gratefully acknowledges the support from Fondecyt Iniciacion Grant Number 111890005. The initial conditions were generated using the AMUSE framework \citep{por09,por13,pel13}. The plots were made with the Veusz plotting package and the Matplolib library \citep{matplotlib}. The simulations were run on the calculation server at the Center for Computational Astrophysics at NAOJ.
	
	\bibliography{ms.bib}
	
\end{document}